\documentclass[twocolumn,english,showpacs,preprintnumbers,amsmath,amssymb,floatfix]{revtex4-1}

\usepackage[T1]{fontenc}
\usepackage[latin9]{inputenc}
\usepackage{color}
\usepackage{array}
\usepackage{amstext}
\usepackage{graphicx}
\usepackage{esint}
\usepackage{rotating}
\usepackage[font=small,labelfont=bf]{caption}
\usepackage{hyperref}

\makeatletter

\@ifundefined{textcolor}{}
{%
 \definecolor{BLACK}{gray}{0}
 \definecolor{WHITE}{gray}{1}
 \definecolor{RED}{rgb}{1,0,0}
 \definecolor{GREEN}{rgb}{0,1,0}
 \definecolor{BLUE}{rgb}{0,0,1}
 \definecolor{CYAN}{cmyk}{1,0,0,0}
 \definecolor{MAGENTA}{cmyk}{0,1,0,0}
 \definecolor{YELLOW}{cmyk}{0,0,1,0}
 }

\@ifundefined{definecolor}
 {\usepackage{color}}{}
\@ifundefined{definecolor}
 {\usepackage{color}}{}
\makeatother
\usepackage{babel}

\begin{document}


\title { Global analysis of nuclear parton distribution functions and their uncertainties at next-to-next-to-leading order }

\author { Hamzeh Khanpour$^{a,b}$ }
\email{Hamzeh.Khanpour@mail.ipm.ir}

\author { S. Atashbar Tehrani$^{c}$ }
\email{Atashbar@ipm.ir}

\affiliation {
$^{(a)}$Department of Physics, University of Science and Technology of Mazandaran, P.O.Box 48518-78195, Behshahr, Iran    \\
$^{(b)}$School of Particles and Accelerators, Institute for Research in Fundamental Sciences (IPM), P.O.Box 19395-5531, Tehran, Iran   \\
$^{(c)}$Independent researcher, P.O.Box 1149-8834413 Tehran, Iran  }

\date{\today}

\begin{abstract}\label{abstract}

We perform a next-to-next-to-leading order (NNLO) analysis of nuclear parton distribution functions (nPDFs) using
neutral current charged-lepton ($\ell ^\pm$ + nucleus) deeply inelastic scattering
(DIS) data and Drell-Yan (DY) cross-section ratios $\sigma_{DY}^{A}/\sigma_{DY}^{A^\prime}$ for several nuclear targets.
We study in detail the parametrizations and the atomic mass (A) dependence of the nuclear PDFs at this order.
The present nuclear PDFs global analysis provides us a complete set of nuclear PDFs, $f_i^{(A,Z)}(x,Q^2)$, with a full functional dependence on $x$, A, Q$^2$.
The uncertainties of the obtained nuclear modification factors for each parton flavour are estimated using the well-known Hessian method. The nuclear charm quark distributions are also added into the analysis.
We compare the parametrization results with the available data and the results of other nuclear PDFs groups. We found our nuclear PDFs to be in reasonably good agreement with them. The estimates of errors provided by our global analysis are rather smaller than those of other groups. In general, a very good agreement is achieved. We also briefly review the recent heavy-ion collisions data including the first experimental data from the LHC proton+lead and lead+lead run which can be used in the global fits of nuclear PDFs. We highlight different aspects of the high luminosity Pb--Pb and p--Pb data which have been recorded by the CMS Collaboration.

\end{abstract}


\maketitle

\tableofcontents{}


\section{Introduction}\label{Introduction}

Deep inelastic scattering (DIS) processes in HERA and hadron collisions in Tevatron and CERN-LHC provide very important tools for probing
the quarks momentum distributions in the nucleons and in the nuclei.
In order to describe the structure of colliding hadrons in DIS processes, a precise knowledge of the parton distribution functions (PDFs) is required.
In order to achieve a better set of PDFs, many groups perform and update their global analyses of PDFs for protons~\cite{Buckley:2014ana,Harland-Lang:2014zoa,Khanpour:2012zz,Khanpour:2012tk,Alekhin:2012ig,Aaron:2009aa,::2014uva,Tung:2006tb,Lai:2010vv,Martin:2009bu,Martin:2010db,Martin:2009iq,Alekhin:2009ni,Gluck:2007ck,CooperSarkar:2010ik,:2009wt,Alekhin:2012du,cteq65,MSTW,JimenezDelgado:2008hf,Owens:2012bv}
and also for nuclei~\cite{Kovarik:2015cma,AtashbarTehrani:2012xh,Kulagin:2014vsa,deFlorian:2011fp,Eskola:2008ca,Kovarik:2010uv,Eskola:2009uj,Hirai:2001np,Hirai:2004wq,Hirai:2007sx,Tehrani:2004hp,Tehrani:2006gy,Tehrani:2007hu,Eskola:1998df,Eskola:1998iy,deFlorian:2003qf,Frankfurt:2011cs,Schienbein:2009kk,deFlorian:2012qw}.
Excellent global fits for the free proton PDFs and nuclear PDFs have been obtained by the mentioned phenomenological groups.
Also a neural network techniques have been successfully developed by NNPDF group~\cite{Ball:2014uwa,Ball:2012cx,Ball:2010de,Ball:2011mu}.
The accuracy of the mentioned PDFs determinations has steadily improved over the recent years, both due to more
accurate DIS data and also due to improvements in perturbation theory predictions for the hard parton scattering reactions.
Since the first indications that the DIS structure functions measured in the charged-lepton scattering of the nuclei, ($\ell ^\pm$ + nucleus), differ
significantly from those measured in the isolated nucleons, there has been also a continuous interest in fully understanding
the microscopic mechanism responsible in the nuclei. The importance of nuclear effects in parton distribution functions is due to the interpretation of any hard-process results involving nuclei in p+A~\cite{Loizides:2013nka}, d+A~\cite{Bozek:2011if} and A+A~\cite{Busch:2015yya,Drozhzhova:2015lla} collisions, such as heavy ions collisions at the present BNL-RHIC~\cite{Dumitru:2001jx,Tannenbaum:2014tea} and CERN-LHC~\cite{Paukkunen:2014nqa} colliders and also future proposed electron-nucleon colliders such as EIC, eRHIC or the LHeC~\cite{Venugopalan:2015jsa,Milner:2014dpa,Aschenauer:2014cki,Alaniz:2014ofa,Alaniz:2014ofa,Boer:2011fh,Polini:2014coa}. These DIS data play an important role for the observed nuclear modifications. The clean experimental environment in the DIS experiment at eRHIC and LHeC would provide a unique opportunity to investigate the nuclear PDFs properties. The nuclear PDFs determination in every QCD analysis have large uncertainties and are not fully constrained by the available DIS data,  consequently further constraints for nuclear PDFs, especially gluons, in such high-energy nuclear colliders in the yet unexplored regions of the $x$ and Q$^2$ plane, are most welcome.

Information about the nuclear PDFs of the nuclei can be extracted from high-energy measurements involving nuclei.
Statistically, most significant data that people use in their nuclear PDFs analysis are the deep inelastic scattering (DIS) experiment which have been taken by experimental groups in fixed-target experiments. These data incorporates: The SLAC (Stanford Linear Accelerator
Center)-E49, E87, E139 and E140 Collaborations, the NMC  (New Muon Collaboration), the EMC (European Muon Collaboration), the BCDMS (Bologna-CERN-Dubna-Munich-Saclay), HERMES, JLAB groups (Jefferson Lab), the Fermilab (Fermi National Accelerator Laboratory)-E665 Collaboration, the Drell-Yan data from the
Fermilab-E772 and E866/NuSea Collaborations~\cite{Amaudruz:1995tq,Gomez:1993ri,Bodek:1983ec,Bodek:1983qn,Dasu:1988ru,Benvenuti:1987az,Ashman:1992kv,
Ashman:1988bf,Adams:1992nf,Adams:1995is,Seely:2009gt,Arneodo:1995cs,Arneodo:1989sy,
Bari:1985ga,Arneodo:1996rv,Arneodo:1996ru,Ackerstaff:1999ac,Vasilev:1999fa,Alde:1990im}. The mentioned data confirm a specific feature of the nuclear reactions called EMC effect at certain region of $x$-Bjorken.
The nuclear PDFs are extracted from global analysis to a wide range of experimental data points. Owing to the complementary nature of the different DIS measurements, tight constraints on the nuclear PDFs can be obtained.  A reliable extraction of nuclear PDFs from the experimental data is required for deeper understanding of the mechanism associated in hard nuclear reactions at RHIC, CERN-LHC and future electron-heavy ion collision. As a result, the kinematic range of data as well as the precise determination of nuclear PDFs will continue to be a topical issue in lots of area of high energy nuclear physics program.

The main difficulties of any global analysis of nuclear PDFs are the lack of precise experimental data points  that we have and fewer types of the data covering kinematical region of $x$ and $Q^2$ which lead to less constraints than the free proton case and also the atomic mass (A) dependence of the nuclear PDFs parameters. Consequently, the nuclear PDFs determination are not simply as the parton densities in the nucleons. In addition still more precise DIS data are needed, especially on the nuclear anti-quark and gluon distributions at very low $x$ to constrain the initial state for the future RHIC and CERN-LHC programs. The DIS data of charged leptons off heavy ion targets are still used in all global nuclear PDFs analyses and provide the best constraints on nuclear modification factors for different parton distributions. These DIS data incorporate a wide range of nuclei from helium to lead which is presented as structure function ratios for different nuclei covered the range of $0.005 \lesssim x \lesssim 1$. The mentioned data will provide enough constraint in obtaining the valence quark distributions. The available data on Drell-Yan (DY) dilepton production of heavy ion target can mainly provide probes a good discrimination between valence and sea quarks distribution in the nuclei. As we mentioned, these type of data only loosely constrain the nuclear modification of gluon distribution due to the limited range in the hard energy scale. As a result, the kinematic range of data as well as the precise determination of nuclear PDFs will continue to be a topical issue in lots of areas of high energy nuclear physics program.

In the present article, we shall present for the first time, a very good quality of the nuclear PDFs using the global analysis of available experimental data, taking into account the ratio of the most commonly analyzed data sets of the structure function ratios, $F_2^{A}/F_2^{A^\prime}$, and Drell-Yan (DY) cross-section ratios $\sigma_{DY}^{pA}/\sigma_{DY}^{pA^\prime}$. Since the first of the nuclear PDFs sets, {\tt AT12}~~\cite{AtashbarTehrani:2012xh}, the procedure has been improved by performing the analysis at the next-to-next-to-leading order (NNLO). An important and appealing feature of the present global QCD analysis of the nuclear PDFs is that we used the theoretical predictions at the next-to-next-to-leading order (NNLO) accuracy in perturbative QCD.
We have performed a careful estimation of the uncertainties using the most common and practical method, the ``Hessian method'' for the nuclear modification factors of the gluons and quarks originating from the experimental errors. The resulting eigenvector sets of the nuclear PDFs can be used to propagate uncertainties to any other desired observable.
The zero-mass variable flavour number scheme (ZM-VFNS) is used in our analysis in order to consider the heavy quarks contributions.

The present nuclear PDFs are characterized by the full functional dependence on $x$, Q$^2$ and atomic mass number (A). We also introduce the additional A dependence directly to the coefficients of the nuclear PDFs at input scale. As in other available nuclear PDFs, we also consider a flavour asymmetric anti-quark distributions. We found no unusual large  uncertainties for nuclear modification factor of the gluon density at medium to large $x$ obtained in some other nuclear PDFs analyses.
Our global analysis considerably leads to smaller value of uncertainties in comparison with other nuclear PDFs global analyses. A detailed comparison with other available nuclear PDFs results including {\tt EPS09}, {\tt HKN07}, {\tt AT12}, {\tt nDS} and {\tt DSSZ12} have been presented.
We also focus on the roles of the NNLO terms on the nuclear PDFs determination by comparing the available NLO results with our NNLO analysis.
The main features of our present NNLO parametrization of nuclear PDFs are worth emphasizing already at this point. It is clear that for a precise nuclear PDFs analysis, more precise data and future advances in the theory will be needed.

The rest of the present paper consists of the following sections. In Sec.~\ref{nPDFs-analysis-method}, we shall provide a formalism to establish an analysis method and a brief discussion of the theoretical structure of the nuclear PDFs, where they arise in the calculation of DIS cross-sections and further theoretical
background relevant to the reliable determination of the nuclear PDFs from experimental data. A brief summary of experimental measurements which are used in the determination of nuclear PDFs is provided in Sec.~\ref{globalnPDFs}.
The analysis method and the error calculation based on the Hessian method are discussed in Sec.~\ref{error-calculation}.
The results of the present nuclear PDFs analysis are given in Sec.~\ref{Results}. In Sec.~\ref{Comparisonwiththedata} a detailed comparison between the present results and available experimental data are presented.  We have attempted a detailed comparison of our NNLO results with recent results from the literature in Sec.~\ref{Comparisonglobalanalyses}. A brief discussion on recent heavy-ion collisions including the first experimental data from the
CERN-LHC proton+lead and lead+lead collisions are presented in Sec.~\ref{LHC-era}. Finally, we have presented our summary and conclusions in Sec.~\ref{Conclusions}. In Appendix A, we present more details on the parameterization and in Appendix B a code is provided for calculating the nuclear PDFs including their uncertainties at given $x$ and Q$^2$ in the NNLO approximation.


\section{Nuclear PDFs analysis method}\label{nPDFs-analysis-method}

In this section, we present our method for global analysis of nuclear PDFs (nPDFs) at next-to-next-to-leading order (NNLO).
In order to calculate the parton distribution in nuclei, we need the parton distributions in a free proton.
We used the following standard parameterizations at the input scale Q$_0^2$=2 GeV$^2$ for all parton species $x q$, obtained from JR09 set of the free proton
PDFs~\cite{JimenezDelgado:2008hf},
\begin{eqnarray}\label{PDFQ0}
x u_v(x,Q_0^2) &=& 3.2350 x^{0.6710} (1 - x)^{3.9293} \nonumber \\
&&(1 - 0.5302 x^{0.5} + 3.9029 x)\nonumber \\
x d_v(x,Q_0^2) &=& 13.058 x^{1.0701} (1 - x)^{6.2177} \nonumber \\
&&(1 - 2.5830 x^{0.5} + 3.8965 x)\nonumber \\
x \Delta (x,Q_0^2) &=& 8.1558 x^{1.328} (1 - x)^{21.043} \nonumber \\
&&(1 - 7.6334 x^{0.5} + 20.054 x)\nonumber \\
x\overline{u}+x\overline{d} &=& 0.4250 x^{-0.1098} (1 - x)^{10.34} \nonumber \\
&&(1 - 3.0946 x^{0.5} + 11.613 x)\nonumber \\
xg &=& 3.0076 x^{0.0637} (1 - x)^{5.54473} \,.
\end{eqnarray}
Here $x u_v$, $x d_v$ represent the valance quark distributions,  $x (\overline{d} + \overline{u})$ the anti-quark distributions, $x\Delta=x(\overline{d}-\overline{u})$, the strange sea distribution $x s = x \overline{s} = \frac{1}{4} x (\overline{d} + \overline{u})$ and the gluon distribution, $x g$.
The nuclear modifications are provided by a number of parameters at a fixed Q$^2$  which are normally denoted by Q$_0^2$. The  nuclear PDFs are
related to the PDFs in a free proton and for this purpose nucleonic PDFs are multiplied by a weight function $w_i(x,A,Z)$.
With the PDFs for a bound proton inside a nucleus A, $f_i(x,Q_0^2)$, one can reconstruct the PDFs for a general nucleus (A, Z) as follow:
\begin{equation}\label{PDFQ0A}
f_i^{(A,Z)} (x,Q_0^2) = w_i(x,A,Z) f^{\rm JR09}_i(x,Q_0^2) \,,
\end{equation}

where $f^{\rm JR09}_i(x,Q_0^2)$ are coming from JR09 parameterization~\cite{JimenezDelgado:2008hf} as they were introduced by Eq.~(\ref{PDFQ0}).
Here we follow  the analysis given by~\cite{AtashbarTehrani:2012xh,Hirai:2007sx,Eskola:2008ca,Rith:2014tma,Eskola:2012rg,Paukkunen:2014nqa}
and assume the following functional form for the nuclear modification as a weight function,
\begin{eqnarray}\label{weight}
 w_i(x, A, Z) & = & 1 + \left( 1 - \frac{1}{A^{\alpha}}\right) \nonumber \\
 && \frac{a_i(A,Z) + b_i(A) x + c_i(A) x^2 + d_i(A) x^3 }{ (1-x)^{\beta_i} }  \,.\nonumber \\
\end{eqnarray}
The parameters in weight function are obtained by a global $\chi^2$ analysis procedure which are dependent on Bjorken variable $x$, mass number A
and atomic number Z.
The important feature of the present analysis is that we let the free parameters of the weight function to have atomic number (A) dependencies.
In order to accommodate various nuclear target materials, we introduce a nuclear A dependence in the weight function coefficients,
\begin{eqnarray}\label{parameters}
b_i(A) \rightarrow b_1 A^{b_2};  \, \, \,  c_i(A) \rightarrow c_1 A^{c_2}; \, \, \, d_i(A) \rightarrow d_1 A^{d_2} \nonumber \\
a_{\overline{q}} (A) \rightarrow   a_1 A^{a_2}      \,.
\end{eqnarray}

Combining the weight function in Eq.~(\ref{weight}) with PDFs of Eq.~(\ref{PDFQ0}), will yield us nuclear PDFs as in what follows:
\begin{eqnarray} \label{partonQW}
u_v^{(A,Z)}(x,Q_0^2) & = & w_{u_v}(x,A,Z)\frac{Z\; u_v(x,Q_0^2)+ N\; d_v(x,Q_0^2)}{A} \nonumber \,,  \\
d_v^{(A,Z)}(x,Q_0^2) & = & w_{d_v}(x,A,Z)\frac{Z\; d_v(x,Q_0^2)+ N\; u_v(x,Q_0^2)}{A} \nonumber \,,  \\
\overline{u}^{(A,Z)}(x,Q_0^2) & = & w_{\overline{q}}(x,A,Z)\frac{Z\; \overline{u}(x,Q_0^2)+ N \, \overline{d}(x,Q_0^2)}{A}\nonumber \,, \\
\overline{d}^{(A,Z)}(x,Q_0^2) & = & w_{\overline{q}}(x,A,Z)\frac{Z\; \overline{d}(x,Q_0^2)+ N \, \overline{u}(x,Q_0^2)}{A}\nonumber \,, \\
s^{(A,Z)}(x,Q_0^2) & = & \overline{s}^{(A,Z)}(x,Q_0^2) = w_{\overline{q}}(x,A,Z)s(x,Q_0^2) \nonumber \,, \\
g^{(A,Z)}(x,Q_0^2) & = & w_{g}(x,A,Z) \, g(x,Q_0^2) \,.
\end{eqnarray}

In the first four equations, the Z term as atomic number indicates the number of protons and the N = A - Z  term represents the
number of neutrons in the nuclei while the SU(3) symmetry is apparently broken there.
We find that the parametrization in Eq.~(\ref{partonQW}) is sufficiently flexible to allow a good $\chi^2$ fit to the available data sets.

If the number of protons and neutrons in a nuclei are equal to each other (iso-scalar nuclei) such as
$^2$D, $^4$He, $^{12}$C and $^{40}$Ca nuclei, the valence quarks $u_v^{(A,Z)}$, $d_v^{(A,Z)}$, $\overline{u}^{(A,Z)}$ and $\overline{d}^{(A,Z)}$
would have similar distributions. In the case that Z and A numbers are not equal in the nuclei, it can be concluded that
anti-quark distributions $\overline{u}^{(A,Z)}$, $\overline{d}^{(A,Z)}$ and $\overline{s}^{(A,Z)}$ in the nuclei would not be equal to each other~\cite{AtashbarTehrani:2012xh,Kumano:1997cy,Garvey:2001yq}.
For the strange quark distributions in the nuclei some research studies are still being done~\cite{Kusina:2012vh} but we assume the common case in which it is assumed $(s = \overline{s})$.

In Eq.~(\ref{weight}) we fixed $\alpha = \frac{1}{3}$, considering the constrains which are imposed by nuclear volume and surface contributions. The parameters $b_i(A)$, $c_i(A)$ and $d_i(A)$ which are listed in Eq.~(\ref{parameters}), will be directly determined from the global $\chi^2$ fits. The fermi motion part parameter $\beta_i$ can not be determined from fit due to the lack of
experimental data. We fixed them to  $\beta_v = 0.4$, $\beta_{\overline{q}} = 0.1$ and $\beta_g = 0.1$ for valence, sea quark and gluon distributions respectively.

There are three constraints for the parameters namely the nuclear charge Z, baryon number (mass number) A and momentum conservations~\cite{Hirai:2001np,Hirai:2004wq,AtashbarTehrani:2012xh,Frankfurt:1990xz},
\begin{eqnarray} \label{Constraints}
 Z & = & \int   \frac{A}{3} \left[2u_v^A - d_v^A\right] (x, Q_0^2)\, dx \,, \nonumber\\
 3 & = & \int   \left[u_v^A + d_v^A\right] (x, Q_0^2) \, dx \,,  \nonumber\\
 1 & = & \int x \left[u_v^A + d_v^A + 2\left\{\overline{u}^A + \overline{d}^A + \overline{s}^A\right\}\right. \nonumber\\
 && \left. + g^A\right] (x,Q_0^2) \, dx \,.
\end{eqnarray}
The $a_i(A,Z)$ parameters for the $u_v$ and $d_v$ distributions ($a_v$) are fixed by the nuclear charge Z and baryon number A conservations, while
$a_g$ parameters for the gluon distribution, is fixed by the existing momentum sum rule in Eq.~(\ref{Constraints}).

\begin{figure}[htb]
\begin{center}
\vspace{1cm}
\resizebox{0.42\textwidth}{!}{\includegraphics{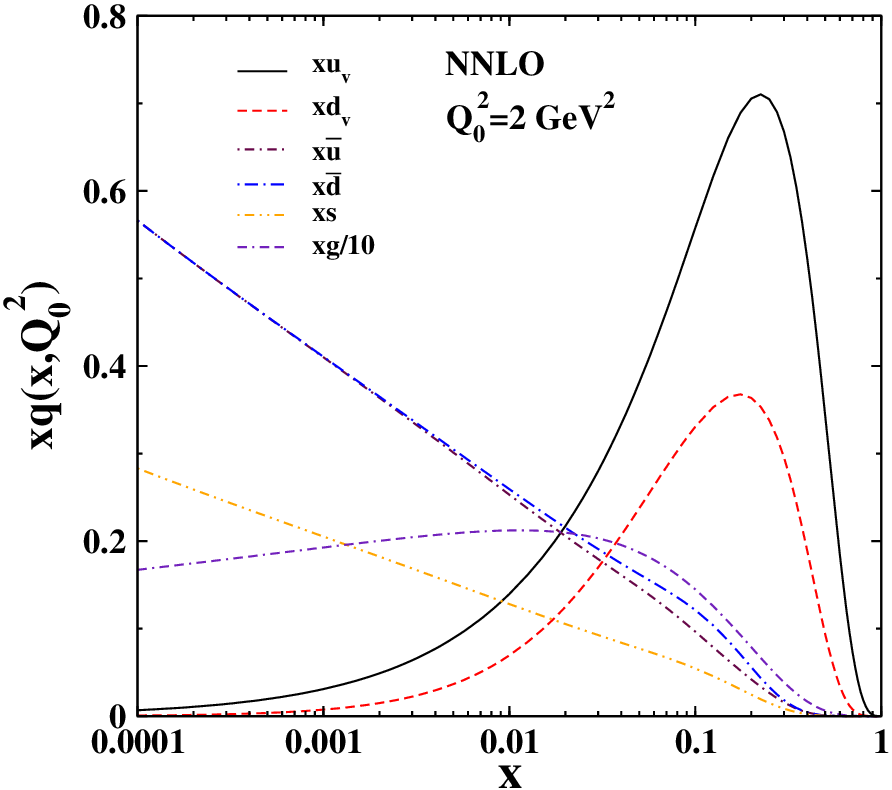}}
\caption{ (Color online) The input PDFs from JR09~\cite{JimenezDelgado:2008hf} at the input scale Q$_0^2$ = 2 GeV$^2$ at NNLO approximation.} \label{JR2009PDF}
\end{center}
\end{figure}

In our calculations, we take Q$_0^2$ = 2 GeV$^2$ and the  $\chi^2$ analysis is done based on the well-known DGLAP evolution
equations~\cite{Vogt:2004ns}. Our calculations are done at the next-next-to-leading (NNLO) approximation in which the modified minimal
subtraction scheme $ ( \overline{MS} ) $  is used.
In our previous next-to-leading order nuclear PDFs analysis~\cite{AtashbarTehrani:2012xh}, the NLO version of the KKT PDFs fit was employed~\cite{Khanpour:2012zz,Khanpour:2012tk}. Since the NNLO version of the mention PDFs fits are not available yet, the JR09 (Jimenez-Delgado, Reya) nucleonic PDFs parametrization~\cite{JimenezDelgado:2008hf} is used in the present analysis. According to their analysis, the strange quark PDFs is assumed to
be symmetric ( $x s = x \overline{s}$ ) and it is proportional to the isoscalar light quark sea and parameterized as
\begin{eqnarray} \label{ssbar}
 x s  = x \overline{s}   = k ( \overline{d} + \overline{u} ) \,,
\end{eqnarray}
with $k$ = $\frac{1}{4}$. In Fig.~\ref{JR2009PDF}, we plot the NNLO parton distribution functions of JR09 at the input scale Q$_0^2$ = 2 GeV$^2$.

For the Q$^2$ evolution and in order to account for the heavy quarks contributions, we choose the zero-mass variable flavour number scheme (ZM-VFNS) with the
charm flavour threshold set at $m_c$ = 1.40 GeV. We add the nuclear charm quark distributions into the present nuclear PDFs analysis.
In the ZM-VFNS, the only explicit dependance on the quark masses in the value at which the number of active flavours changes.
We let the heavy quarks to be massless and generate them through the DGLAP evolution above the mass thresholds.

The $F_2^{(A,Z)} (x,Q^2)$ structure functions can be extracted at NNLO approximation as a convolutions of nuclear PDFs of
Eq.~(\ref{PDFQ0A}) with the corresponding Wilson coefficients~\cite{Vermaseren:2005qc,vanNeerven:1999ca,vanNeerven:2000uj},
\begin{eqnarray} \label{F2A2}
 F_2^{(A,Z)} (x,Q^2)  = \sum_{i = u, d, s, g} C_i \otimes f_i^{(A,Z)} (x,Q^2) \,.
\end{eqnarray}
Consequently the nuclear structure functions are given by
\begin{eqnarray} \label{F2A23}
 F_2^{(A,Z)}(x,Q^2)  =
 && \sum_{i = u, d, s} e_i^2 x\left[ 1 +  a_sC^1_{q}(x) + a_s^2 C^2_{q}(x)\right]  \nonumber \\
 && \otimes ( q_i^A + \overline{q}_i^A )   \,. \nonumber \\
 && + \frac{1}{2f}(a_s C^1_g(x) + a_s^2 C^2_g(x)) \otimes xg \,.
\end{eqnarray}
In this equation, $C^{1, 2}_{q, g}$ are the common Wilson coefficient at NLO and NNLO approximation~\cite{vanNeerven:1999ca,vanNeerven:2000uj} and the symbol $\otimes$ denotes the usual convolution integral,
\begin{equation}\label{covoloution}
f(x)\otimes g(x) = \int_x^1\frac{dy}{y}f\left(\frac{x} {y}\right)g(y) \,.
\end{equation}


\section{Input to the global nuclear PDFs fit}\label{globalnPDFs}

In the present section, we review the available experimental data including charged-lepton ($\ell ^\pm$ + nucleus) DIS
and Drell-Yan cross-section ratios $\sigma_{DY}^{pA}/\sigma_{DY}^{pA^\prime}$ for different nuclear targets as the input for the global fit.
In order to include the heavy-target data into a global analysis of proton PDFs, the nuclear corrections are considered.
Using these variety of $\ell ^\pm$A and Drell-Yan data, we can construct global nuclear PDFs fit.
A large and complete experimental data sets for different nuclear targets in wide range of $x$ and Q$^2$ required to fully
constraints the $x$, A, Q$^2$ and also for flavour dependencies of the nuclear PDFs.
The nuclear effects have been studied experimentally in charged lepton-nucleus scattering
by some experimental groups such as the muon experiments BCDMS, EMC and NMC at CERN, EMC-NA38 and E665 at FNAL, in electron scattering at
SLAC, DESY and JLAB, in the Drell-Yan process and also in neutrino-nucleus scattering.

\begin{figure}[htb]
\begin{center}
\vspace{1cm}
\resizebox{0.42\textwidth}{!}{\includegraphics{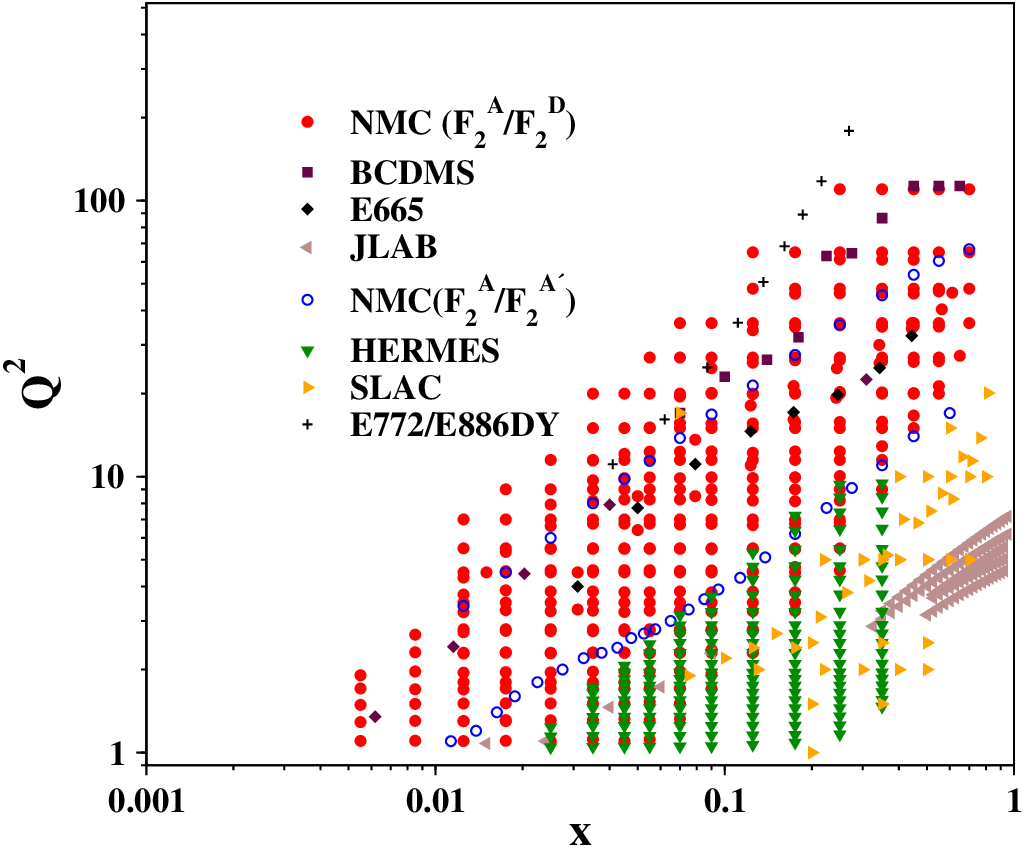}}   
\caption{ (Color online) Nominal coverage of the data sets used in our global fits. The plot nicely summarizes the universal $x$ dependance of the nuclear effect.}\label{figxQdata}
\end{center}
\end{figure}

The $x$ and Q$^2$ coverage of the data sets used in our nuclear PDFs fits are illustrated in figure~\ref{figxQdata}.
The interval range of the Q$^2$ values is Q$^2 \geq 1$ GeV$^2$ and the smallest value for the $x$-Bjorken variable
is equal to 0.0055 at this stage. Nominally there is a substantial amount of data at larger Bjorken variable $x$. The plot clearly shows the worse coverage of the data at medium to small-$x$ region.

As figure~\ref{figxQdata} clearly shows, these data sets are latter limited in comparison with the data for free proton PDFs.
The proton fit uses a very large and very precise data from  Tevatron and HERA colliders while the nuclear PDFs uses a smaller data sample
from several fixed target experiments and some collider data from RHIC.
Consequently, lacking precision and smaller amount of the nuclear data specially at small value of $x$-Bjorken, could lead to larger uncertainties for
nuclear PDFs than the PDFs for the free proton. Consequently for better and precise determination of nuclear quark and gluon distributions, especially for very low
parton momentum fractions $x$, further measurements for the EMC effect in the neutrino-nucleus, electron-nucleus and proton-nucleus scattering are needed.

The total experimental data sets that we used in our present global analysis are listed in Table~\ref{tabledata}.
The $F_2^A$ ($F_2^{A^\prime}$) is denoting the structure function of a nuclei and $F_2^D$ is representing the structure function of deuterium.
Number of data points, together with the related references and specific nuclear targets are also listed in the table.
The total number of data sets for the $F_2^A/F_2^D$ ratios are equal to 1079 and the number of $F_2^A/F_2^{A^{\prime}}$ ratio for Be/C, Al/C, Ca/C, Fe/C, Sn/C, Pb/C, C/Li is 308.
The data comes from the Drell-Yan process provide a complementary constraint on the nuclear PDFs. In particular, they allow one to separate the sea quark distributions in the nuclei. For this purpose, we use the data obtained by FNAL-E866~\cite{Vasilev:1999fa} and FNAL-E772~\cite{Alde:1990im} experiments at Fermilab.
For the Drell-Yan cross section ratios ($\sigma_{DY}^{pA}/\sigma_{DY}^{pA^\prime}$) we have 92 data points while the related ratio are C/D,  Ca/D,  Fe/D,  W/D,  Fe/Be  and  W/Be.
The total  experimental data points were included in our analysis is 1479. They contain  lepton-nucleus deep inelastic scattering ($\ell ^\pm$ + nucleus)  and Drell-Yan cross-section ratios $\sigma_{DY}^{pA}/\sigma_{DY}^{pA^\prime}$ data for different nuclear targets.

\begin{table*}
\small
\begin{center}
\begin{tabular}{ c c c c}
\hline\hline Nucleus     &     Experiment   &   Number of data points   &   Reference    \\
\hline\hline
  F$_{2}^{A}$/F$_{2}^{D} $   &     &    &   \\  \hline
He/D & SLAC-E139   & 18 &       \cite{Arneodo:1989sy}     \\
& NMC-95           & 17 &       \cite{Amaudruz:1995tq}    \\
Li/D & NMC-95      & 17 &       \cite{Amaudruz:1995tq}    \\
Li/D(Q$^{2}$dep.)  & NMC-95 & 179 &   \cite{Arneodo:1995cs}   \\
Be/D & SLAC-E139   & 17 &       \cite{Gomez:1993ri}     \\
C/D & EMC-88       & 9 &        \cite{Ashman:1988bf}    \\
& EMC-90           & 5 &        \cite{Arneodo:1989sy}   \\
& SLAC-E139        & 7 &        \cite{Gomez:1993ri}     \\
& NMC-95           & 17 &       \cite{Amaudruz:1995tq}   \\
& FNAL-E665        & 5 &        \cite{Adams:1995is}      \\
& JLAB-E03-103     & 103 &      \cite{Seely:2009gt}      \\
C/D(Q$^{2}$dep.) & NMC-95 & 191 &   \cite{Arneodo:1995cs}    \\
N/D & BCDMS-85     & 9 &        \cite{Bari:1985ga}      \\
& HERMES-03        & 153 &      \cite{Ackerstaff:1999ac}    \\
Al/D & SLAC-E49    & 18 &       \cite{Bodek:1983ec}      \\
& SLAC-E139        & 17 &       \cite{Gomez:1993ri}      \\
Ca/D & EMC-90      & 5 &        \cite{Arneodo:1989sy}    \\
& NMC-95           & 16 &       \cite{Amaudruz:1995tq}  \\
& SLAC-E139        & 7 &        \cite{Arneodo:1989sy}   \\
& FNAL-E665        & 5 &        \cite{Adams:1995is}     \\
Fe/D & SLAC-E87    & 14 &       \cite{Bodek:1983qn}     \\
& SLAC-E139        & 23 &       \cite{Gomez:1993ri}     \\
& SLAC-E140        & 10 &       \cite{Dasu:1988ru}      \\
& BCDMS-87         & 10 &       \cite{Benvenuti:1987az}   \\
Cu/D & EMC-93      & 19 &       \cite{Ashman:1992kv}      \\
Kr/D & HERMES-03   & 144 &      \cite{Ackerstaff:1999ac}    \\
Ag/D & SLAC-E139   & 7 &        \cite{Gomez:1993ri}     \\
Sn/D & EMC-88      & 8 &         \cite{Ashman:1988bf}   \\
Xe/D & FNAL-E665-92 & 5 &        \cite{Adams:1992nf}     \\
Au/D & SLAC-E139   & 18 &        \cite{Gomez:1993ri}     \\
& SLAC-E140        & 1 &         \cite{Dasu:1988ru}     \\
Pb/D & FNAL-E665-95 & 5 &        \cite{Adams:1995is}     \\ \hline\hline
 F$_{2}^{A}$/F$_{2}^{A^{\prime }}$   &     &    &        \\  \hline
Be/C & NMC-96     & 15 &         \cite{Arneodo:1996rv}    \\
Al/C & NMC-96     & 15 &         \cite{Arneodo:1996rv}    \\
Ca/C & NMC-96     & 24 &         \cite{Amaudruz:1995tq}   \\
& NMC-96          & 15 &          \cite{Arneodo:1996rv}   \\
Fe/C & NMC-96     & 15 &         \cite{Arneodo:1996rv}    \\
Sn/C & NMC-96     & 146 &        \cite{Arneodo:1996rv}   \\
     & NMC-96     & 15  &        \cite{Arneodo:1996ru}   \\
Pb/C & NMC-96     & 15 &        \cite{Arneodo:1996rv}    \\
C/Li & NMC-95     & 24 &        \cite{Amaudruz:1995tq}   \\
Ca/Li & NMC-95    & 24 &        \cite{Amaudruz:1995tq}     \\ \hline\hline
$ \sigma _{DY}^{A}/\sigma _{DY}^{A^{\prime }} $  &   &   &    \\ \hline
Fe/Be & FNAL-E866/NuSea   & 28 &    \cite{Vasilev:1999fa}    \\
W/Be & FNAL-E866/NuSea    & 28 &    \cite{Vasilev:1999fa}    \\
C/D & FNAL-E772-90   & 9 &      \cite{Alde:1990im}    \\
Ca/D & FNAL-E772-90  & 9 &      \cite{Alde:1990im}    \\
Fe/D & FNAL-E772-90  & 9 &      \cite{Alde:1990im}    \\
W/D & FNAL-E772-90   & 9 &      \cite{Alde:1990im}    \\ \hline \hline
Total &                   & 1479 &  \\   \hline
\end{tabular}
\caption[]{ The charged-lepton DIS experimental data sets for F$_{2}^{A}$/F$_{2}^{D}$, F$_{2}^{A}$/F$_{2}^{A^{\prime}}$ and Drell-Yan cross section ratios $\sigma _{DY}^{A}/\sigma _{DY}^{A^{\prime }}$ used in the present global fit. Number of data points, the related references and specific nuclear targets are also listed. }
\label{tabledata}
\end{center}
\end{table*}


\section{The analysis of $\chi^2$ value and error calculation via Hessian method}\label{error-calculation}

To determine the best fit at NNLO, we need to minimize the $\chi^2$ with respect to 16 free input nuclear PDFs parameters of Eq.~(\ref{partonQW}).
The global goodness-of-fit procedure follows the usual chi--squared method with $\chi^2 (p)$ defined as

\begin{equation} \label{chi}
\chi^2 (p) = \sum_{i=1}^{n^{\rm data}} \frac{ ( R_i^{\rm data} - R_i^{\rm theory} (p) )^2}{ ( \sigma_i^{\rm data} )^2 } \,,
\end{equation}
where $p$ denotes the set of 16 independent parameters in the fit and $n^{data}$ is the number of
data points included, $n^{data}$ = 1479 for the NNLO fit.
The optimization of the above $\chi ^2$ value to determine the best parametrization of the nuclear PDFs is done by the CERN program library {\tt MINUIT}~\cite{MINUIT}.
For the $i^{\mathrm{th}}$ experiment, $R_{i}^{\rm data}$, $\sigma_i^{\rm data}$, and $R_{i}^{\rm theory}$
denote the experimental data value, measured uncertainty and theoretical value for the $n^{\mathrm{th}}$ data point.
The experimental errors are calculated from systematic and statistical errors, $\sigma_i^{\rm data} = \sqrt{ (\sigma_i^{\rm sys})^2 + (\sigma_i^{\rm stat})^2 }$.
The theory prediction $R_{i}^{\rm theory}$, which is denoting the theoretical result of $F_{2}^{p A}$/$F_{2}^{p A^{\prime }}$ and $\sigma _{ DY }^{pA}/\sigma_{ DY }^{pA^{\prime }}$ ratios, depends on the input nuclear PDFs parameters $p$.

\begin{figure}[htb]
\begin{center}
\vspace{1cm}
\resizebox{0.40\textwidth}{!}{\includegraphics{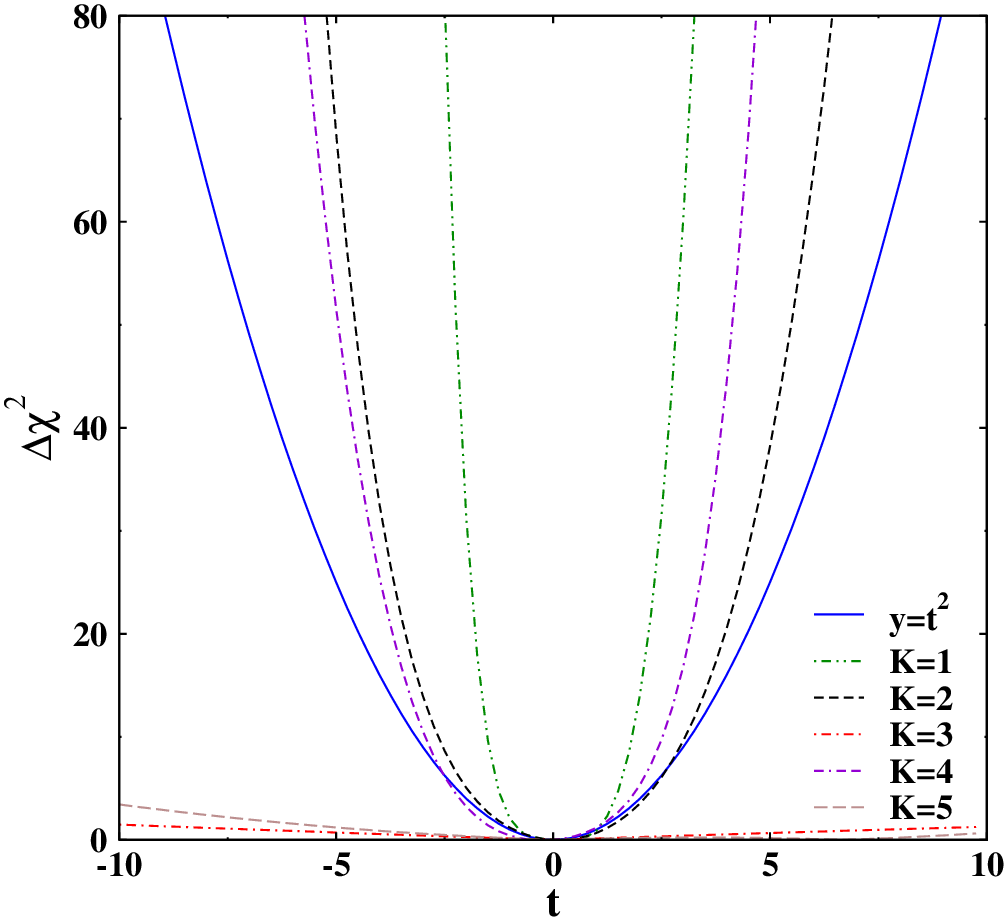}}  
\caption[]{ (Color online) $\Delta \chi^2$ as a function of $t$ defined in Eq.(\ref{pi}) for some random sample
of eigenvectors.}\label{fig:deltachi}
\end{center}
\end{figure}

For the error calculation, a standard error analysis is needed for the nuclear PDFs by taking into account correlations among the parameters.
The method to consider the correlations among the uncertainties are discussed in details in Refs.~\cite{AtashbarTehrani:2012xh,Martin:2002aw,Martin:2009iq,Pumplin:2001ct,Monfared:2011xf,Arbabifar:2013tma}, so we explain only a brief outline here.
Following that, an error analysis can be done using the Hessian or covariance matrix, which is obtained by running the CERN program library {\tt MINUIT}.
The nuclear PDFs uncertainties are estimated, using the Hessian matrix as the following
\begin{eqnarray}\label{hessian}
 \delta f^A(x)&=&\nonumber\\
 \biggl[\Delta \chi^2 &&\sum_{i,j}  \left( \frac{\partial f^A(x,\xi)}{\partial \xi_i}
 \right) _{\xi = \hat{\xi}}  H_{ij}^{-1}  \left( \frac{\partial f^A(x,\xi)}{\partial \xi_j} \right)_{\xi = \hat{\xi}} \biggr]^{1/2},\nonumber\\
\end{eqnarray}
where the $H_{ij}$ is the Hessian matrix (also known as the error matrix), $\xi_i$ is the quantity referring to the parameters which exist in nuclear PDFs and $\hat\xi$ indicates the number of parameters which make an extremum value for the related derivative.
We are able to calculate the nuclear PDFs uncertainties using these covariance matrix elements based on the method as mentioned in this section.
Their values at higher Q$^2$ > Q$_0^2$ are calculated by the well-known DGLAP evolution equations.

The well-known Hessian method which is based on the covariance matrix diagonalization, provides us a
simple and efficient method for calculating the PDFs uncertainty~\cite{AtashbarTehrani:2012xh,Martin:2009iq,Pumplin:2001ct,Monfared:2011xf,Arbabifar:2013tma}.
In this method, one can assume that the deviation in the global goodness-of-fit quantity, $\Delta \chi^2_{\rm global}$, is quadratic in the deviation of the parameters specifying the input parton distributions, $p_i$, from their values at the minimum, $p_i^{\rm min}$.
So one can write
\begin{equation}\label{Deltachi}
 \Delta \chi _{\rm global}^{2} \equiv  \chi _{\rm global}^{2}-\chi _{\rm min }^{2} = \sum_{i,j}H_{ij}(p_{i}-p_{i}^{\rm min })(p_{j}-p_{j}^{\rm min }) \,,
\end{equation}
where $H_{i,j}$ is an element of the Hessian matrix determined in the global nuclear PDFs fit.
By having a set of appropriate nuclear PDFs fit parameters which minimize the global $\chi^{2}$ function, $s^{\rm min}$, and introducing nuclear parton sets $s^{\pm}_k$, one can write
\begin{equation}\label{pi}
 p_{i}(s_{k}^{\pm })=p_{i}(s^{\rm min })\pm t\sqrt{\lambda _{k}}v_{ik} \,,
\end{equation}
where $v_{ik}$ is the eigenvector and $\lambda _{k}$ is the $k^{\rm th}$ eigenvalue.
The parameter $t$ is adjusted to make the required $ T = \sqrt{\Delta \chi _{\rm global}^{2}}$ global which is the allowed deterioration in $\Delta \chi _{\rm global}^{2}$
global quality for the error determination and $t$ = $T$ is the ideal quadratic behaviour.
To test the quadratic approximation of Eq.~(\ref{Deltachi}), we study the dependence
of $\Delta \chi _{\rm global}^{2}$ along some random samples of eigenvector directions.
The $\Delta \chi _{\rm global}^{2}$ treatment for some selected eigenvectors numbered $k$ = 1, 2, 3, 4 and 5 for the presented nuclear PDFs analysis are illustrated
in figure~\ref{fig:deltachi}.


\begin{figure}[htb]
	\begin{center}
		\vspace{1cm}
		\resizebox{0.45\textwidth}{!}{\includegraphics{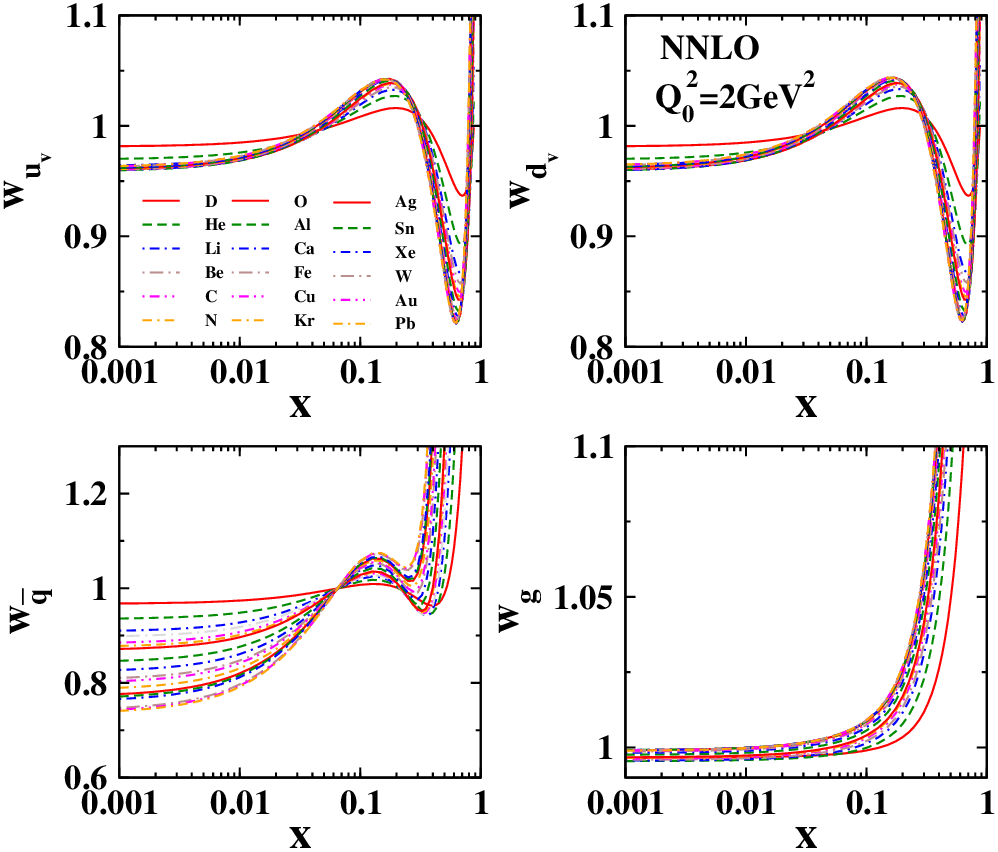}}  
		\caption{(Color online) Nuclear modification factors for $W_{u_v}, W_{d_v}, W_{\overline{q}}$ and $W_g$ are shown in the NNLO for all the analyzed nuclei at
			$Q_0^2$ = 2 GeV$^2$. The nuclear mass number becomes larger in the order of D, He, Li, Be, $ \cdots $, and Pb.}\label{fig:wnnlo}
	\end{center}
\end{figure}

\section{Results of the nuclear PDFs fits}\label{Results}

We are now in a position to present the results of our nuclear PDFs analysis which we call the {\tt KA15} nPDFs fit.
In the following section, the results of the present nuclear PDFs studies are discussed in details and compared with the available experimental data.
In the present analysis which has been done at the NNLO approximation, we obtain an overall  $\frac{\chi^2} {d.o.f} = 1696.65/1463 = 1.15$. The total number of the data points
for the nuclear structure functions and Drell-Yan ratios is 1479. As we mentioned, the number of parameters which is used in our fitting procedure is equal to 16.
The output of the global fit is the set of $b_i$, $c_i$ and $d_i$ parameters which are  corresponding to $b (A)$, $c (A)$ and $d (A)$. Their A-dependent functions will lead to the determination of the nuclear PDFs at the initial scale Q$_0^2$ = 2 GeV$^2$,~ $f_i(x,Q_0^2)$.
At the first step, 20 parameters have been optimized by minimizing the usual chi--squared method $\chi^2 (p)$ and in the second step since we have fixed four parameters $\beta_v = 0.4$, $\beta_{\overline{q}} = 0.1$, $\beta_g = 0.1$ and $\alpha = \frac{1}{3}$, we just need to determine 16 parameters of the weight functions via our
fitting procedure. In order to control the fermi motions of the partons
inside the nuclei at the larger values of $x$, we have to fix $\beta_v$, $\beta_{\overline{q}}$ and $\beta_g$ parameters. These parameters can not be well determined from fit due to the lack of the
experimental data. Consequently fixing these data may lead to reach a well converging (well constrained) global nuclear PDFs fit. For the nuclear modification of the valance and sea quark distributions, we choose an
$A$-dependent functional form while the weight function for the gluon distribution is assumed to be independent of the A number. The numerical
values of the parameters defining the modifications as well as the fixed
parameters are listed in Table.~\ref{table2}. The parameters $a_{u_v}$, $a_{d_v}$ and $a_g$ are fixed by the three
sum rules, given by Eq.~(\ref{Constraints}) (See appendix A of Ref.~\cite{AtashbarTehrani:2012xh} for more details.).
The parameter errors quoted in the table, are due to the propagation of the systematic and statistical errors in the used DIS data.

\begin{table*}
\small
\begin{center}
\begin{tabular}{lll}
\hline  \hline
$a_{v}$      &     $a_{\overline{q}} (A)$     &     $a_{g}$    \\
Appendix A    &    $-0.14364\pm 8.938466\times10^{-3} A^{0.149757\pm 1.3456148\times10^{-2}}$ &   Appendix A      \\
$b_{v} (A)$    &     $b_{\overline{q}} (A)$    &    $b_{g}$     \\
$1.98347\pm 0.1705875 A^{-0.0791784\pm 1.19181\times10^{-2}}$ & $3.1188\pm 0.2080143 A^{0.159521\pm 1.4907795\times10^{-2}}$ & $ 0.105397\pm 2.139654$  \\
$c_{v} (A)$      &     $c_{\overline{q}} (A)$    &   $c_{g}$    \\
$-6.46451\pm 0.3582447 A^{-0.038812\pm1.36899\times10^{-2}}$ & $-15.5991\pm 1.1211789 A^{0.183694\pm 1.8131510\times10^{-2}}$    &   $0 $  \\
$d_{v} (A)$   &   $d_{\overline{q}} (A)$      &   $d_{g}$     \\
$4.90165\pm 0.3045687 A^{0.00900608\pm 1.81409\times10^{-2}}$ & $18.7266\pm 2.2757606 A^{0.255328\pm 2.9314540\times10^{-2}}$   &   $1.48382\pm  1.353835$   \\
$\beta _{v}$   &   $\beta _{\overline{q}}$   &   $\beta _{g}$   \\
$0.4$ Fixed    &   $0.1$ Fixed     &    $0.1$  Fixed   \\ \hline  \hline
\end{tabular}
\caption[]{The input nuclear PDFs parameters of valance quark, sea quark and gluon distributions at Q$_0^2$ = 2 GeV$^2$ obtained by global $\chi^2$ analysis. The details
of the $\chi^2$ analysis and constraints applied to control the parameters are contained in the text.}\label{table2}
\end{center}
\end{table*}

Nuclear modifications $W_{i}$ ($i = u_v$, $d_v$, $\overline{q}$ and $g$) for all the analyzed nuclei at the input scale Q$_0^2$ = 2 GeV$^2$
have been represented in figure~\ref{fig:wnnlo}.

\begin{figure}[htb]
	\begin{center}
		\vspace{1cm}
		\resizebox{0.45\textwidth}{!}{\includegraphics{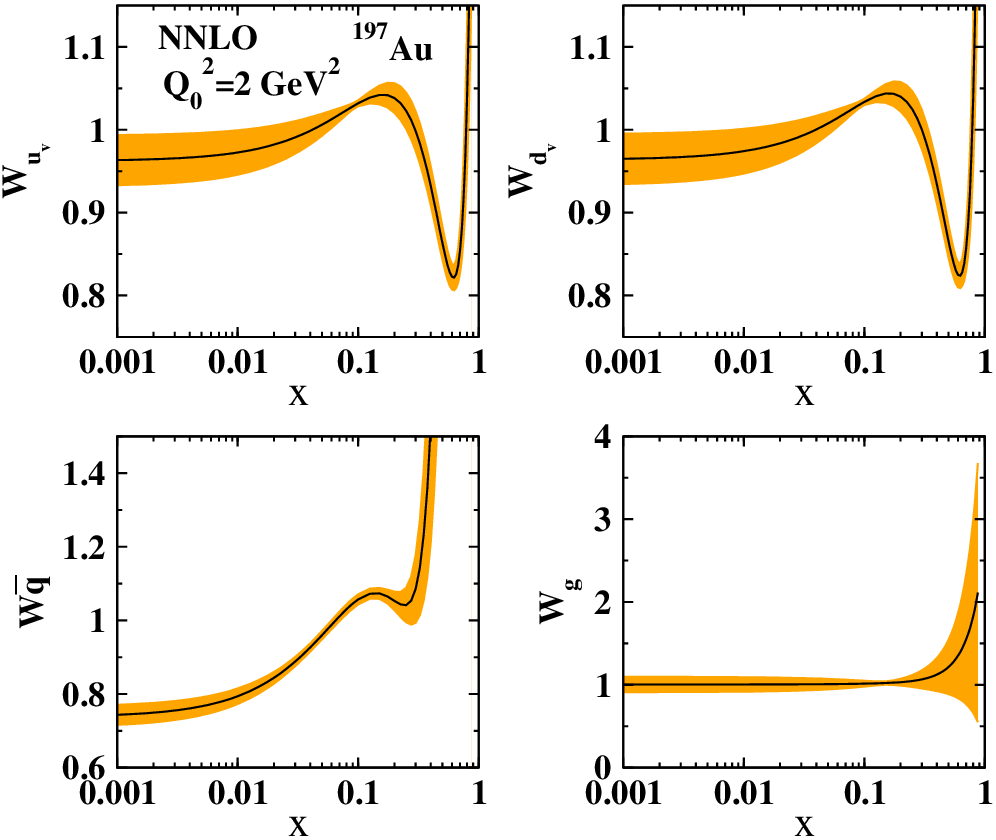}}  
		\caption{(Color online) Nuclear modification factors of the PDFs and their uncertainties are shown for the Gold nucleus at Q$_0^2$ = 2 GeV$^2$.}\label{fig:wnnloAu}
	\end{center}
\end{figure}

\begin{figure}[htb]
	\begin{center}
		\vspace{1cm}
		\resizebox{0.45\textwidth}{!}{\includegraphics{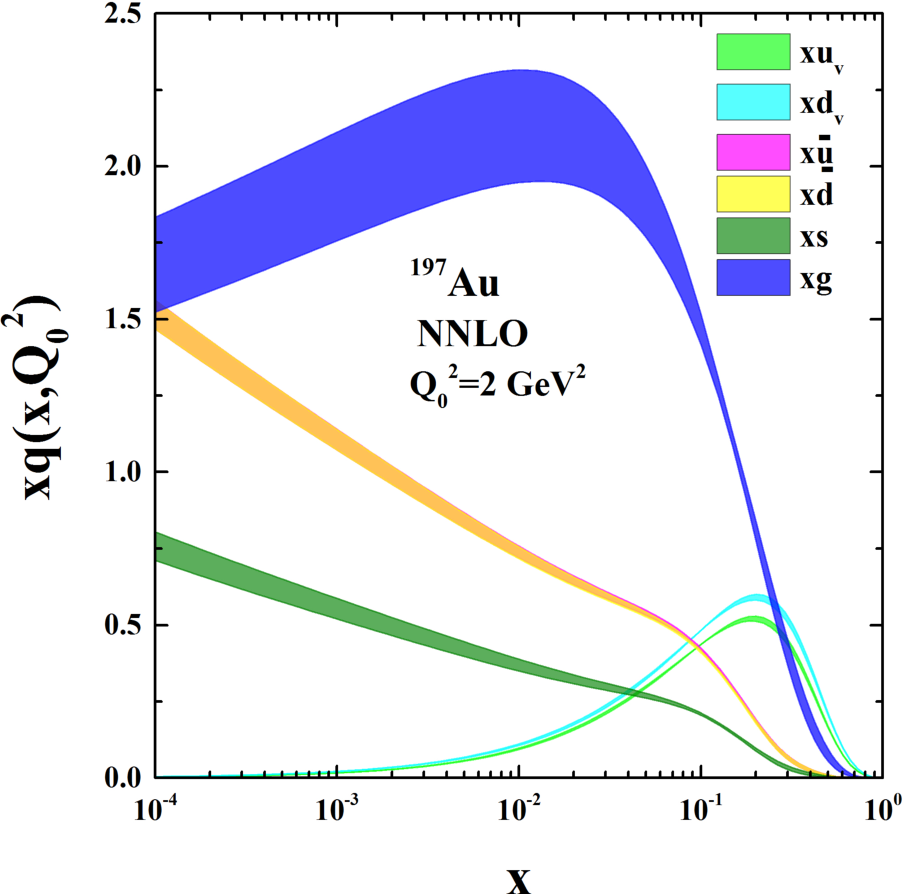}}  
		\caption{(Color online) The nuclear parton distribution functions in Gold nucleus at Q$_0^2$ = 2 GeV$^2$ including their uncertainties. The gluon has a large uncertainty at small-$x$.}\label{fig:AuQ0}
	\end{center}
\end{figure}

As a typical heavy-sized nucleus, gold nuclei has been selected for showing the nuclear modifications
in figure~\ref{fig:wnnloAu} and the nuclear PDFs including their uncertainties in figure~\ref{fig:AuQ0} at the input scale Q$_0^2$ = 2 GeV$^2$.


It is worth noting that although the nuclear PDFs
for $x\overline{d}$ and $x\overline{u}$ are similar, there is small difference with $x\overline{d}$ > $x\overline{u}$. This maybe explained by a relative
suppression of the $g \rightarrow u \bar {u}$ process due to the exclusion principle and the larger number of up quark which already occupied.
The gluon modification and its distribution which plotted for Gold nucleus in figure~\ref{fig:AuQ0} clearly show that we have a large uncertainty band at small value of the Bjorken-$x$ value. The wide uncertainty band for the gluon reflects the fact that there are no enough data constraints.

\begin{figure}[htb]
	\begin{center}
		\vspace{1cm}
		\resizebox{0.42\textwidth}{!}{\includegraphics{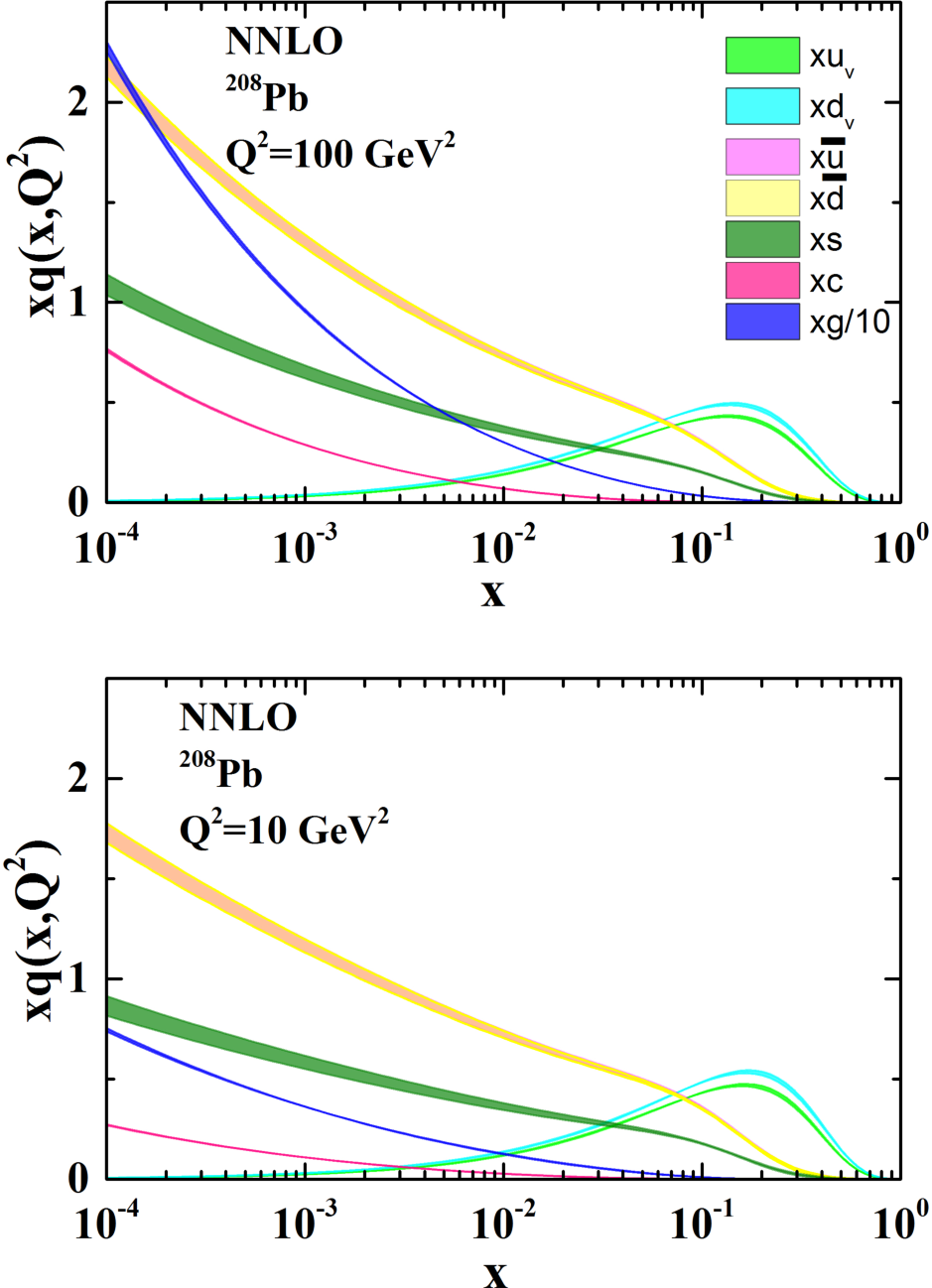}}  
		\caption{(Color online) Nuclear parton distribution functions in Lead at Q$^2$ = 10 GeV$^2$ and 100 GeV$^2$ at the NNLO approximation.}\label{fig:pb}
	\end{center}
\end{figure}

\begin{figure}[htb]
	\begin{center}
		\vspace{1cm}
		\resizebox{0.42\textwidth}{!}{\includegraphics{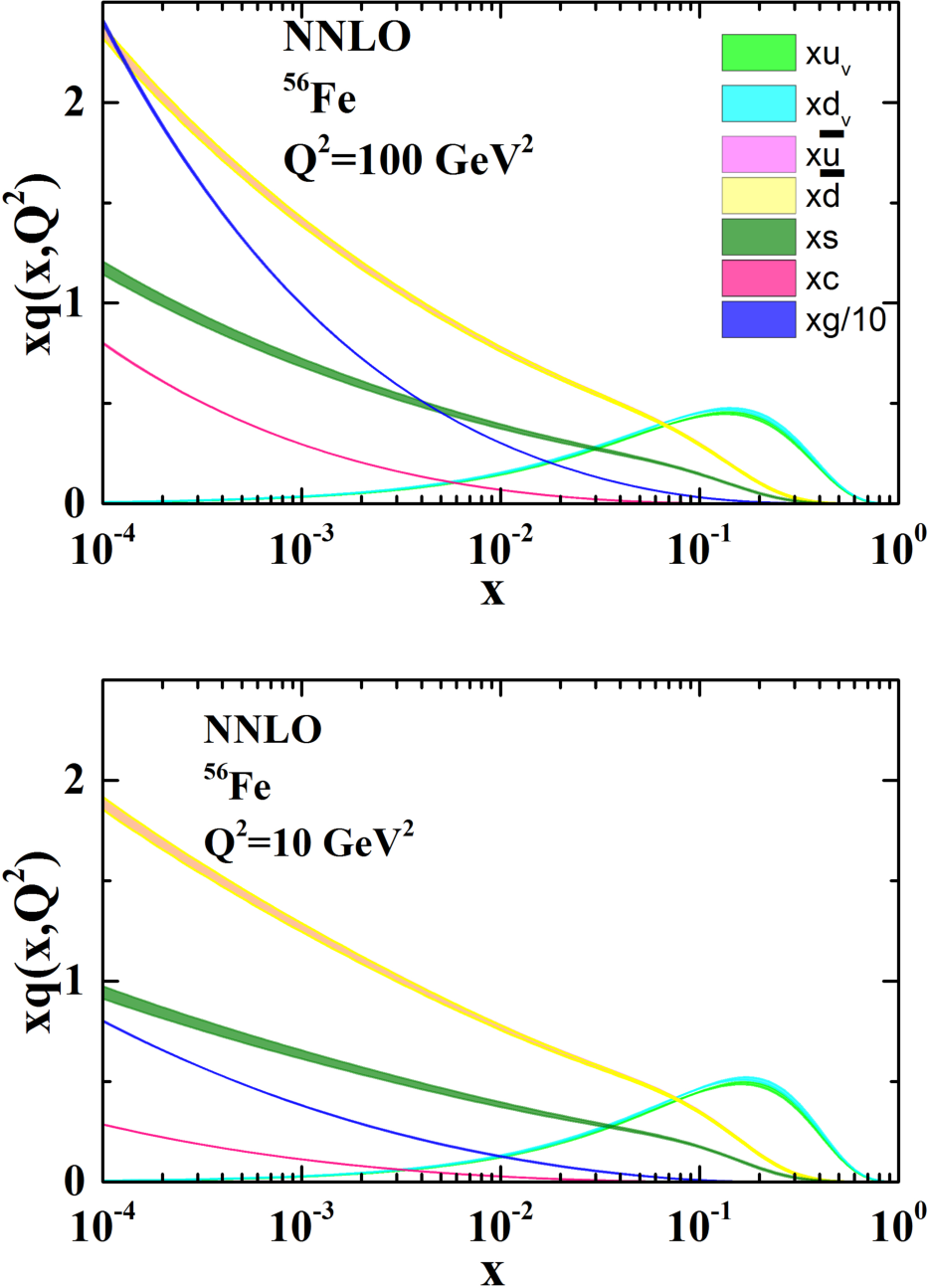}}  
		\caption{(Color online) Nuclear parton distribution functions in Iron at $Q^2$ = 10 GeV$^2$ and 100 GeV$^2$ at the NNLO approximation.}\label{fig:fe}
	\end{center}
\end{figure}

Using the DGLAP evolution equations, we can evolve the $f_i(x,Q_0^2)$ to an arbitrary Q$^2$ to obtain the desired nuclear PDFs $f_i(x,Q^2)$.
In figures~\ref{fig:pb} and ~\ref{fig:fe}, we display the nuclear PDFs at the Q$^2$ =10 GeV$^2$ and 100 GeV$^2$ as a function of $x$ for Lead and Iron respectively.


The $x u_v$, $x d_v$ valance quark distributions, the anti-quark distributions $x\overline{d}$ and $x\overline{u}$, the strange sea distribution $x s$ and also
charm distribution $xc$ and the gluon distribution, $x g$ are shown as well.
As the results clearly show, there are still large uncertainties in the nuclear PDFs, especially for the gluon sectors.
To resolve the gluon uncertainties in nuclei at small-$x$, ($x < 0.001$), much accurate hard scattering data from electron-A collider would be needed.
Further DIS data from RHIC $d$ + Au and CERN-LHC proton lead collisions, will help in constraining the nuclear PDFs.
As we defined in Eq.~\ref{partonQW}, we assumed flavour asymmetric anti-quark distributions, ${\overline d}^A \neq {\overline u}^A$.
In the isoscalar nuclei such as $^2$D, $^4$He, $^{12}$C and $^{40}$Ca, the ${\overline u}^A$ and ${\overline d}$ distributions are equal so we have flavour symmetry.
For other nuclei which the number of their protons and neutrons are not equal, we have the SU(3) flavor symmetry breaking.

\begin{figure}[htb]
	\begin{center}
		\vspace{0.95cm}
		\resizebox{0.4\textwidth}{!}{\includegraphics{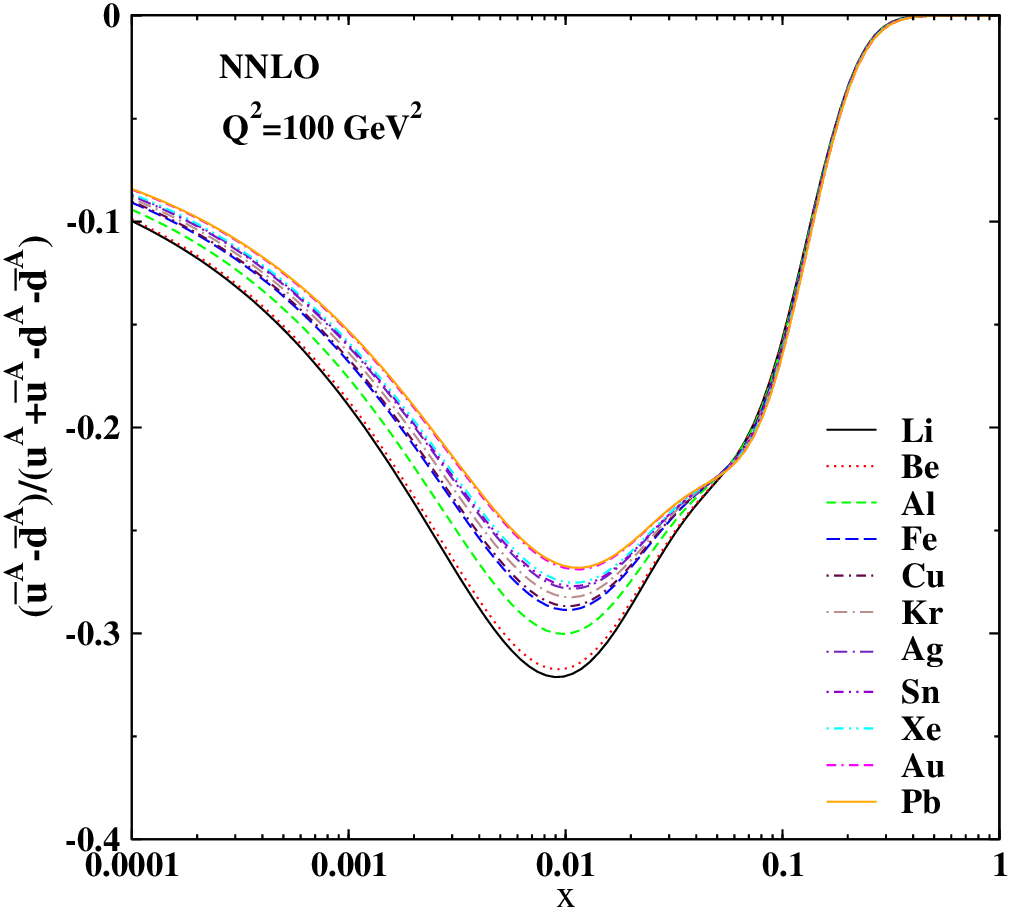}} 
		\caption{(Color online) The ratio of flavour asymmetric distributions, $(\overline{u}^A - \overline{d}^A)/(u^A + \overline{u}^A - d^A - \overline{d}^A)$ is
			shown for some nuclei that have experimental data at Q$^2$ = 100 GeV$^2$. In the isoscalar nuclei, the distributions vanish,
			$( \overline{d}^A - \overline{u}^A = 0)$.}\label{fig:ubar-dbar}
	\end{center}
\end{figure}

Figure~\ref{fig:ubar-dbar} shows a very interesting results. In this figure, the $(\overline{u}^A - \overline{d}^A)/(u^A + \overline{u}^A - d^A -\overline{d}^A)$ ratio has been shown at Q$^2$ = 100 GeV$^2$ for some non-isoscalar nuclei which we have SU(3) symmetry breaking. The ${\overline d}^A \neq {\overline u}^A$ asymmetry are clearly shown at small $x$, ($x < 0.05$). This effect may be due to the sensitivity of the Drell-Yan data to the isospin asymmetry of the sea quark distributions.

\begin{figure}[htb]
	\begin{center}
		\vspace{1cm}
		\resizebox{0.4\textwidth}{!}{\includegraphics{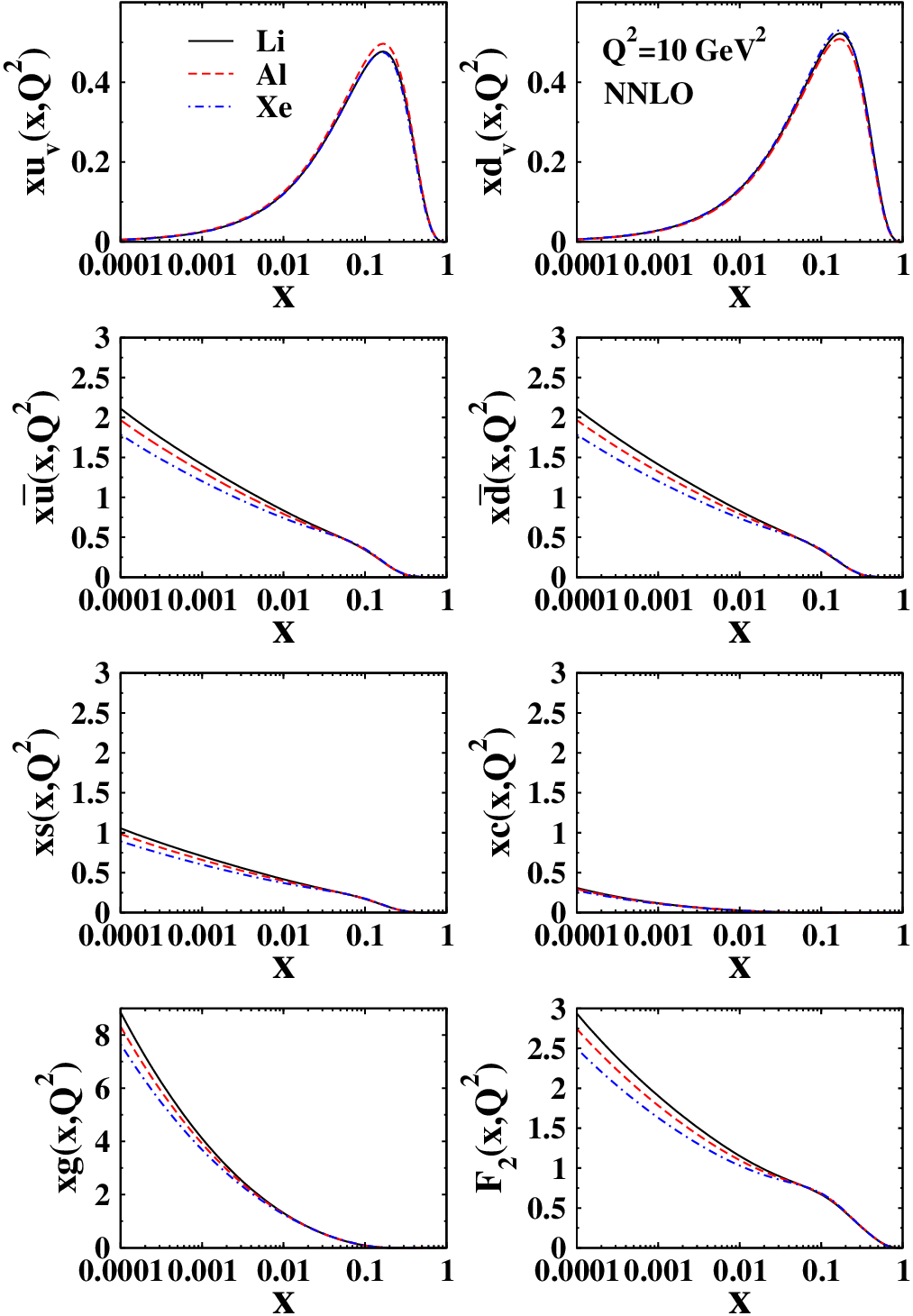}}     
		\caption[]{(Color online) Comparison between nuclear parton distributions in NNLO for a range of nuclei such as Li, Al and Xe nucleus at the Q$^2$ = 10 GeV$^2$. }\label{fig:nuclon}
	\end{center}
\end{figure}

In figure~\ref{fig:nuclon}, we compare nuclear parton distributions of the Li, Al and Xe nucleus at the Q$^2$ = 10 GeV$^2$ to investigate the A-dependence of the various nuclear PDFs flavours. As the plot shows, when examining the A-dependence of nuclear PDFs, we notice that the smaller nuclei has a larger value of sea-quark and gluon distributions at small value of $x$. The $\bar{d}$ and $\bar{u}$ PDFs are very similar because we directly determine the $\bar{d} + \bar{u}$ combination from the analysis.

%
%
\section{Comparison with the experimental data} \label{Comparisonwiththedata}

\begin{figure}[htb]
	\begin{center}
		\vspace{10mm}
		\resizebox{0.45\textwidth}{!}{\includegraphics{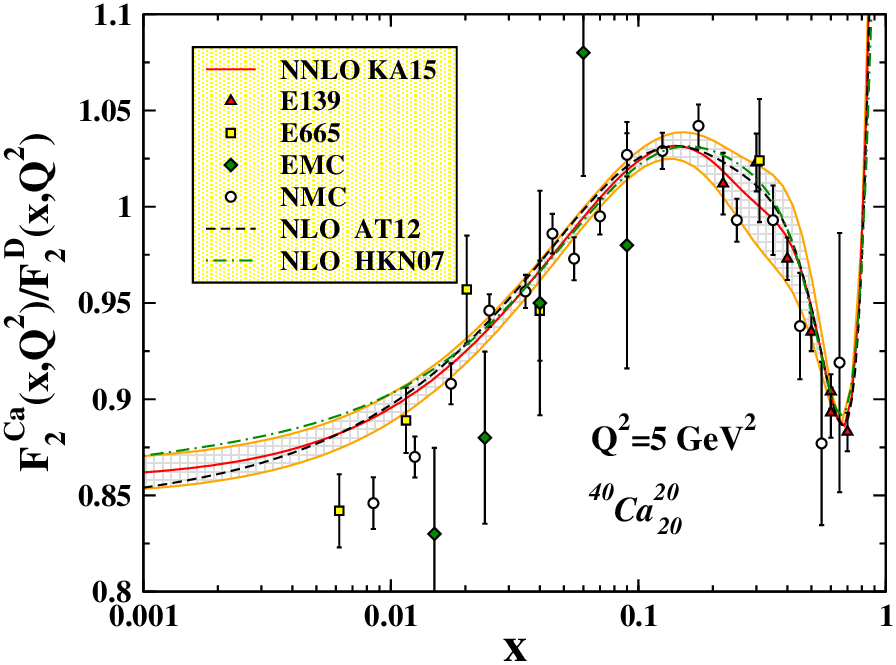}}  
		\caption{(Color online) EMC effect for Calcium nucleus at Q$^2$=5 GeV$^2$ in NNLO approximation and its comparison with our previous
			NLO analysis \cite{AtashbarTehrani:2012xh}. The results from HKN07~\cite{Hirai:2007sx} are also have been shown. In this plot, theoretical results are compared with the
			$F_2^{A (=Ca)}/F_2^{D}$ data. The uncertainties are shown by the shaded bands.}\label{fig:CaQ5}
	\end{center}
\end{figure}

A detailed comparison of our NNLO nuclear PDFs results with the experimental used in this analysis is presented in this section. The error bars in the figures correspond to the statistical and systematic errors added in quadrature.
As we mentioned, the available data are taken in the limited $x$ range without small $x$ data, which leads to difficulty in determining the
nuclear gluon distribution. Figure~\ref{fig:CaQ5} shows this issue. In this figure, we plot the theoretical prediction including uncertainties for structure function ratio of
Calcium nucleus, $F_2^{A (= Ca)}/F_2^{D}$, which has been compared with actual data at Q$^2$=5 GeV$^2$.
Our previous next-to-leading order nuclear PDFs analysis {\tt AT12}~\cite{AtashbarTehrani:2012xh} and the results from {\tt HKN07}~\cite{Hirai:2007sx} are also shown as well.
The theoretical predictions are shown by the curves in the figure and the uncertainties are shown by the shaded bands.
The plot shows that our NNLO parametrizations are successful in explaining the $x$ dependence of the Calcium data as an example.

\begin{figure}[htb]
	\begin{center}
		\vspace{0.95cm}
		\resizebox{0.45\textwidth}{!}{\includegraphics{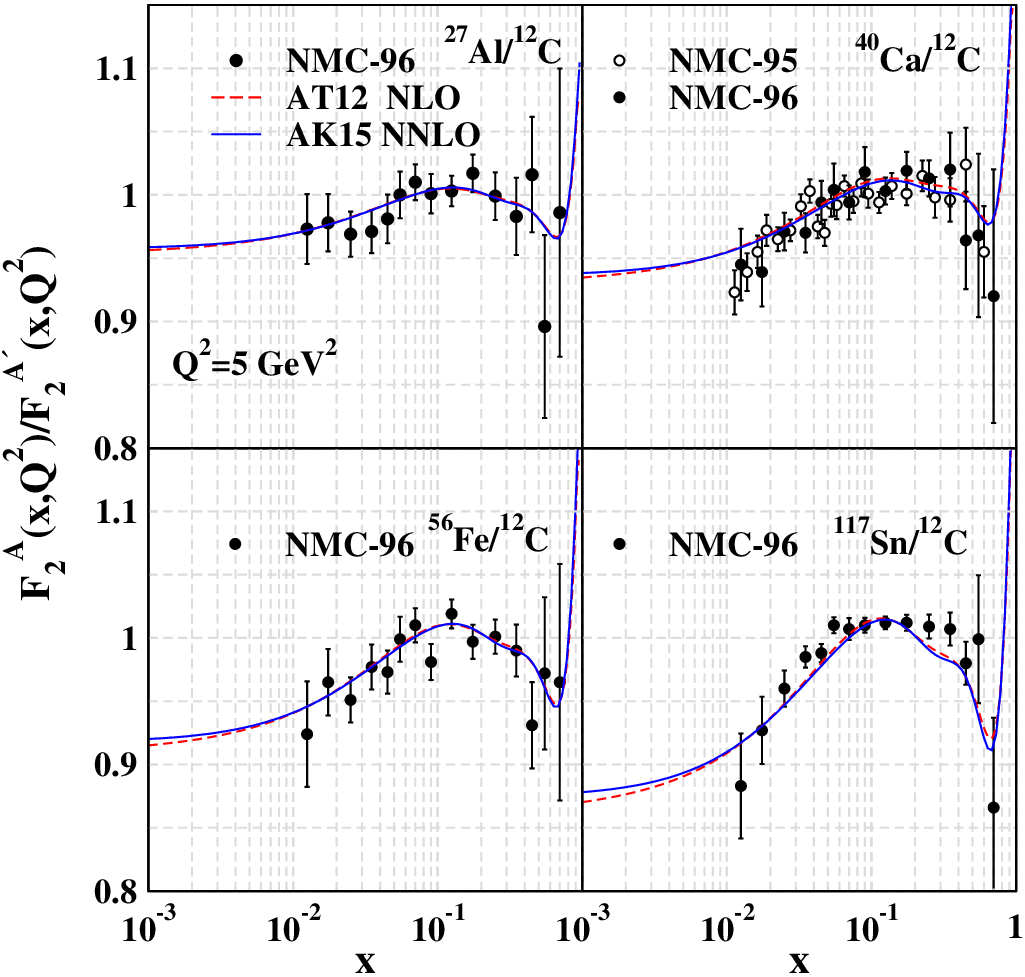}} 
		\caption{(Color online) Comparison of the {\tt KA15} NNLO and {\tt AT12} NLO theory predictions for $R = F_2^A(x,Q^2)/F_2^{A^{\prime}}(x,Q^2)$ as a function of $x$ at the scale Q$^2$=5 Ge$V^2$. The data from NMC has been shown for comparison.}\label{fig:emc4fig}
	\end{center}
\end{figure}

Figure~.~\ref{fig:emc4fig} shows the ratio $R=F_2^A(x,Q^2)/F_2^{A^{\prime}}(x,Q^2)$ in comparison to NMC data for a variety of nuclear targets. The plot clearly shows that both {\tt KA15} NNLO and {\tt AT12} NLO theory predictions describe the data well.

\begin{figure}[htb]
	\begin{center}
		\vspace{1cm}
		\resizebox{0.45\textwidth}{!}{\includegraphics{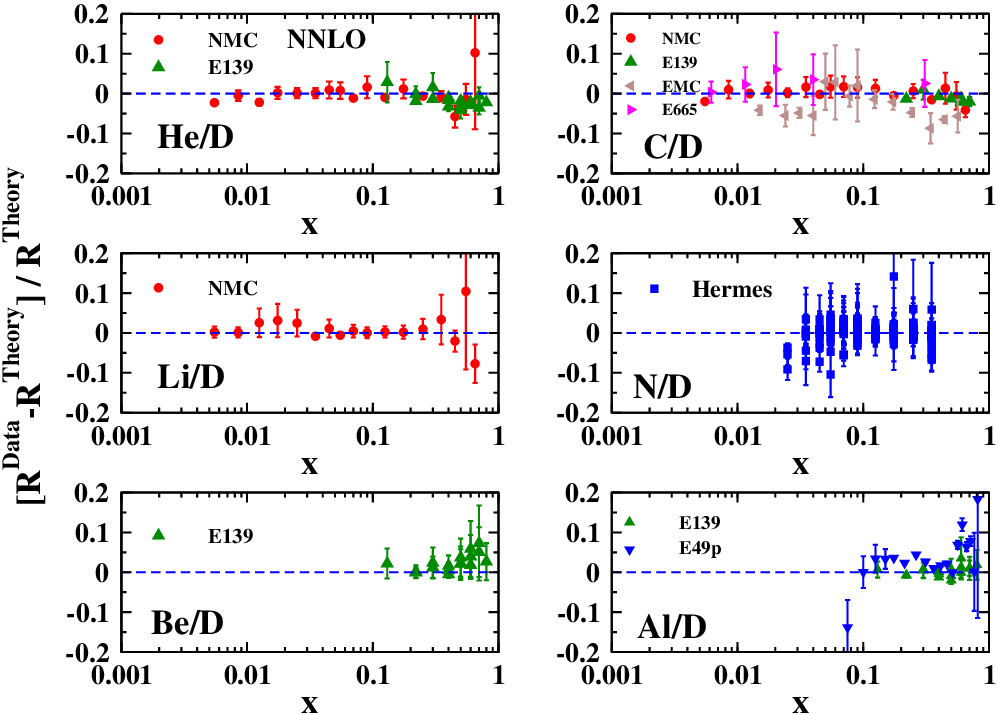}}  
		\caption[]{(Color online) Comparison with the experimental data of $R = F_2^A/F_2^{D}$. The ratios of $(R^{\rm data} - R^{\rm theory})/R^{\rm theory}$ are shown for comparison. The NNLO parametrization is used for the theoretical calculations at the $Q^2$ points of the
			experimental data.} \label{fig:emc1}
	\end{center}
\end{figure}

\begin{figure}[htb]
	\begin{center}
		\vspace{1cm}
		\resizebox{0.45\textwidth}{!}{\includegraphics{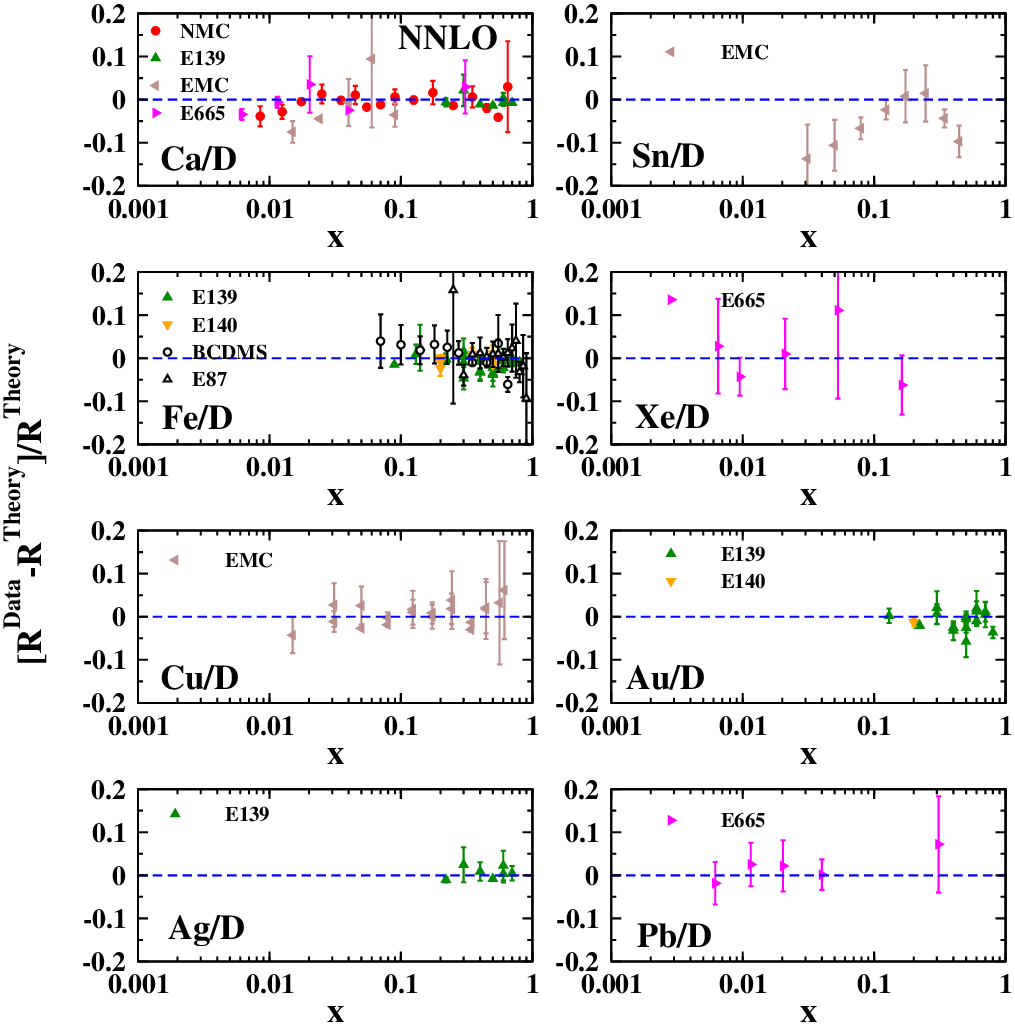}}  
		\caption[]{(Color online) Comparison with experimental data of $R = F_2^A/F_2^{D}$. The ratios of $(R^{\rm data} - R^{\rm theory})/R^{\rm theory}$ are shown for comparison.
			The NNLO parametrization is used for the theoretical calculations at the $Q^2$ points of the experimental data. }\label{fig:emc2}
	\end{center}
\end{figure}

A detailed comparison with the experimental data of the structure function ratios $R = F_2^A/F_2^{D}$ for the analyzed nuclei are shown in figures~\ref{fig:emc1} and \ref{fig:emc2}.


\begin{figure}[htb]
	\begin{center}
		\vspace{1cm}
		\resizebox{0.45\textwidth}{!}{\includegraphics{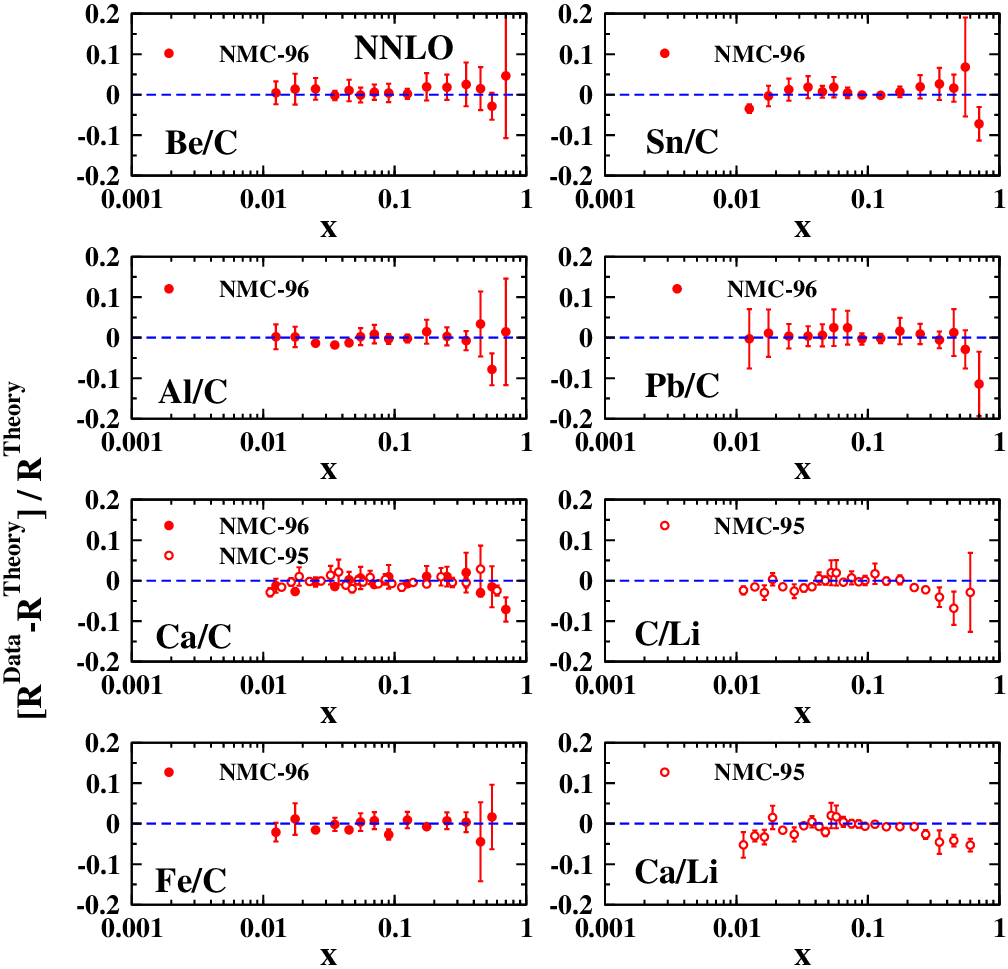}}
		\caption[]{ (Color online) Comparison with experimental data of $R = F_2^A/F_2^{C}$ and $R = F_2^A/F_2^{Li}$. The ratios of $(R^{\rm data} - R^{\rm theory})/R^{\rm theory}$ are shown for comparison. The NNLO parametrization is used for the theoretical calculations at the $Q^2$ points of the
			experimental data.}\label{fig:emc3}
	\end{center}
\end{figure}

The ratios of $(R^{\rm data} - R^{\rm theory})/R^{\rm theory}$ are also shown for comparison. $R^{\rm data}$ is the experimental value and $R^{\rm theory}$ is the theoretical value of the structure function ratios.
The same plots for the structure function ratios of $R = F_2^A/F_2^C$ and $R = F_2^A/F_2^{Li}$ are also have been shown in figure~\ref{fig:emc3}.
The comparison indicates that our NNLO parametrizations should be successful in explaining the $x$ dependance of the analyzed nuclei experimental data.

\begin{figure}[htb]
	\begin{center}
		\vspace{10mm}
		\resizebox{0.4\textwidth}{!}{\includegraphics{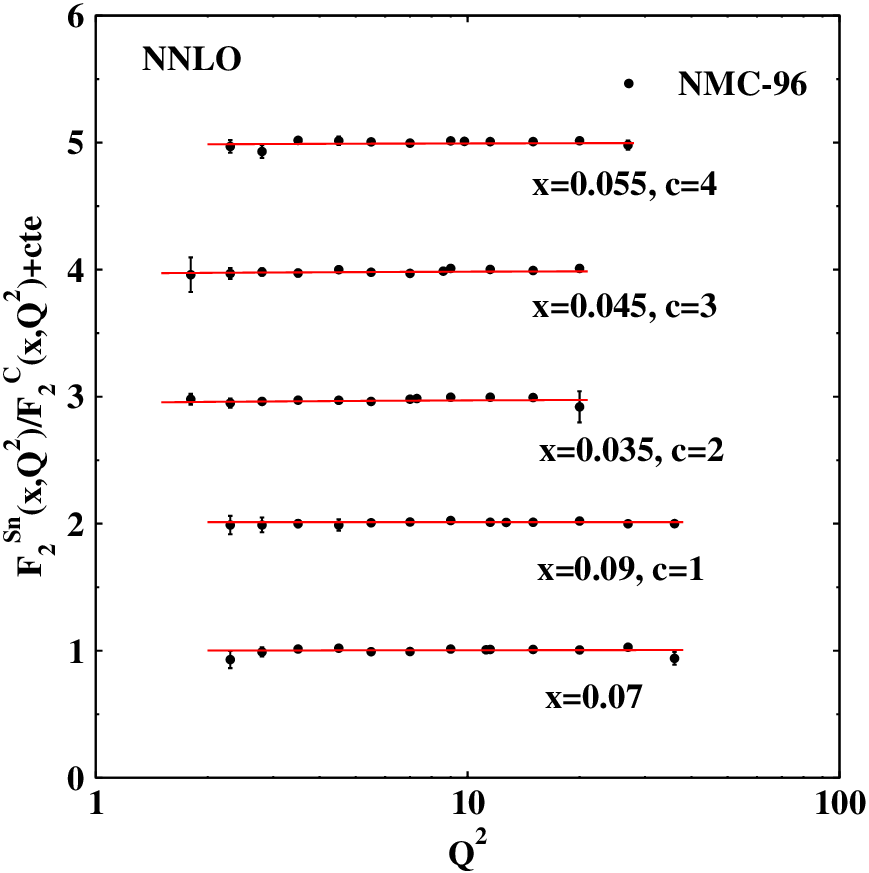}}  
		\caption{ (Color online) The Q$^2$ dependence of the structure function ratios $F_2^{Sn}/F_2^C$ at NNLO in comparison with the experimental data of NMC-96.} \label{fig:sntocQ2}
	\end{center}
\end{figure}

In order to better investigation of the nuclear PDFs, we plot the Q$^2$ dependence of the structure function ratios $F_2^{Sn}/F_2^C$ at NNLO in comparison with the experimental data of NMC-96 in figure~\ref{fig:sntocQ2}. The comparison are shown for some selected smaller values of $x$, $x$ = 0.07, 0.09, 0.035, 0.045 and 0.055.
The results indicating the overall Q$^2$ dependencies are in very good agreements with the data.

\begin{figure}[htb]
	\begin{center}
		\vspace{10mm}
		\resizebox{0.45\textwidth}{!}{\includegraphics{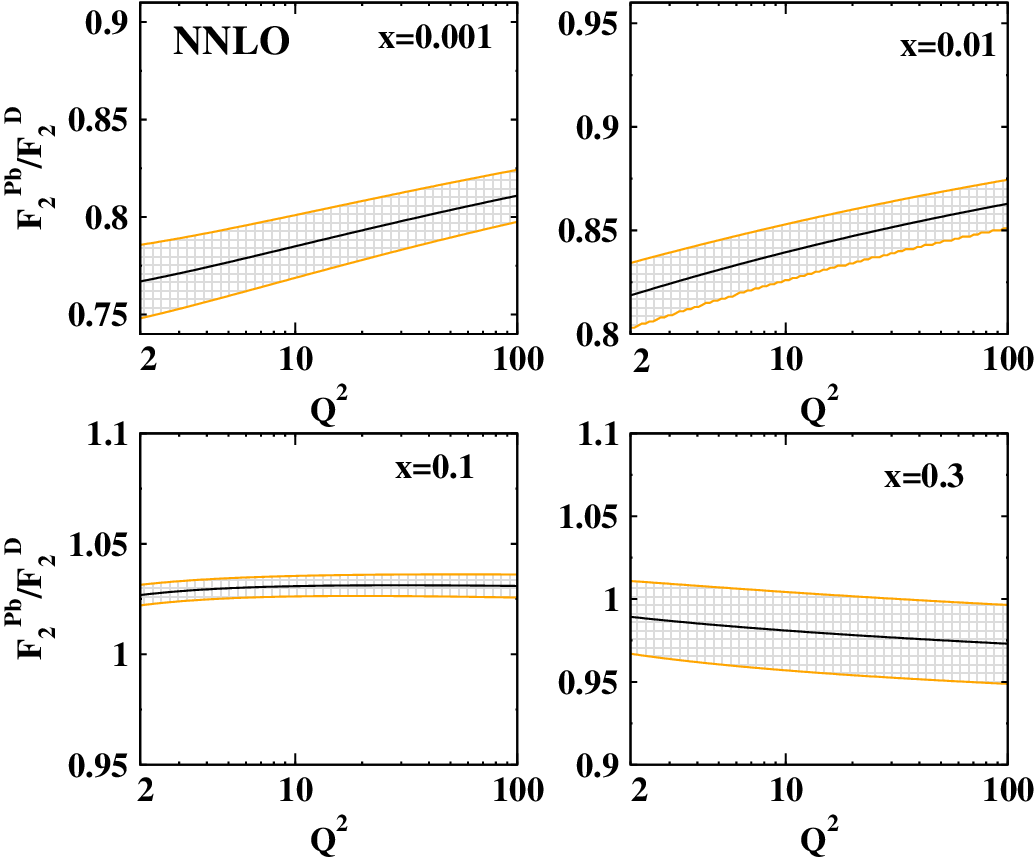}}  
		\caption{ (Color online) $Q^2$ dependence of the theoretical predictions of the structure function ratios $F_2^{Pb} / F_2^D$ at NNLO for different value of  $x$ = 0.001, 0.01, 0.01 and 0.3 including their uncertainties.} \label{fig:F2pbF2DQ}
	\end{center}
\end{figure}

Q$^2$ dependence of the theoretical predictions of the structure function ratios $F_2^{Pb} / F_2^D$ at NNLO for different value of $x$,  $x$ = 0.001, 0.01, 0.01 and 0.3 including their uncertainties have been shown in figure~\ref{fig:F2pbF2DQ}. The theoretical predictions are shown by the curves in the figure and the uncertainties are shown by the shaded bands.

\begin{figure}[htb]
	\begin{center}
		\vspace{0.95cm}
		\resizebox{0.40\textwidth}{!}{\includegraphics{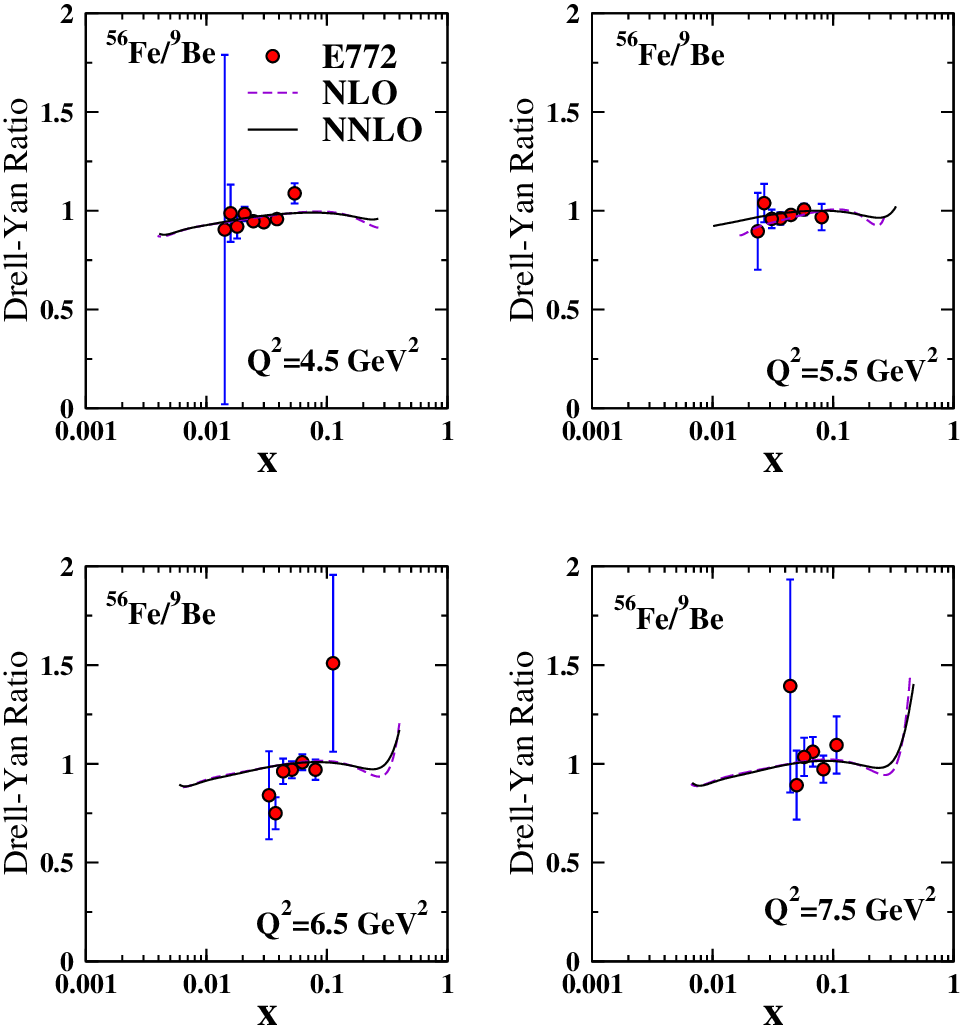}} 
		\caption{ (Color online) Theoretical predictions are compared with the data of the Drell-Yan cross-section ratios $\sigma_{DY}^{Fe}/\sigma_{DY}^{Be}$. Data points are from the FNAL-E866~\cite{Vasilev:1999fa} experiments at Fermilab.}\label{fig:DYFEtoBe}
	\end{center}
\end{figure}

Using the Drell-Yan data of proton-nucleus scattering, one can investigate the nuclear modification of anti-quark distributions.
In figure~\ref{fig:DYFEtoBe}, the theoretical predictions are compared with the data of the Drell-Yan cross-section ratios $\sigma_{DY}^{Fe}/\sigma_{DY}^{Be}$ measured by
FNAL-E866~\cite{Vasilev:1999fa} at Q$^2$ = 4.5 GeV$^2$, 5.5 GeV$^2$, 6.5 GeV$^2$ and 7.5 GeV$^2$. Our previous results for nuclear PDFs at NLO are also shown as well.

The FNAL-E866 data on Drell-Yan cross-section are in a good agreement with the {\tt AT12} NLO and {\tt KA15} NNLO predictions. The cross section of the Drell-Yan process is to small to study any process with colliding ion beams at higher center-of-mass energies. As we mentioned, the data from proton-nucleus and proton-deuteron collisions at the CERN-LHC or RHIC would be very desirable in order to determine the nuclear PDFs at low values of parton fractional momenta $x$~\cite{Accardi:2004be,Salgado:2011wc}.

%
%
\section{Comparison with different global analyses of nuclear PDFs} \label{Comparisonglobalanalyses}

We now in position to compare our NNLO nuclear PDFs {\tt KA15} with other recent nuclear parton distributions in the literature.
Specifically, we will compare our results with the following nuclear PDFs sets: {\tt AT12}~\cite{AtashbarTehrani:2012xh}, {\tt EPS09}~\cite{Eskola:2009uj}, {\tt HKN07}~\cite{Hirai:2007sx}, {\tt nDS}~\cite{deFlorian:2003qf} and {\tt DSSZ12}~\cite{deFlorian:2012qw}. We will briefly summarize the key development of the most recent ones of these.
The initial scale in {\tt EPS09}~\cite{Eskola:2009uj} is set to Q$_0$ = 1.3 GeV and it uses the CTEQ6.1M free proton NLO PDFs. The ZM-VFNS heavy-quark scheme are adopted and the data from $\ell + A$ DIS and p + A DY and $\pi^0$ production in d + Au collisions at PHENIX are used in the {\tt EPS09} nuclear PDFs analysis.   
The {\tt DSSZ12}~\cite{deFlorian:2012qw} uses the free proton NLO PDFs of MSTW, consequently the nuclear modification factors are parameterize at Q$_0$ = 1 GeV. Heavy quarks effects are included using general-mass variable-flavour number scheme (GM-VFNS). This analysis covers the most extensive selection of the nuclear data including $\ell^\pm$-DIS data, p + A DY data together with $\nu$-DIS and $\pi^0$ production in d + Au collisions from PHENIX and STAR. The latest {\tt HKN07} global analysis of nuclear PDFs presented in~\cite{Hirai:2007sx} uses the MRST98 NLO parametrization for the nucleonic PDFs. This analysis covers the  $\ell + A$ DIS and p + A DY data. Charm quark contributions are included and the strange quark contributions are assumed to be symmetric.

\begin{figure}[htb]
	\begin{center}
		\vspace{0.95cm}
		\resizebox{0.45\textwidth}{!}{\includegraphics{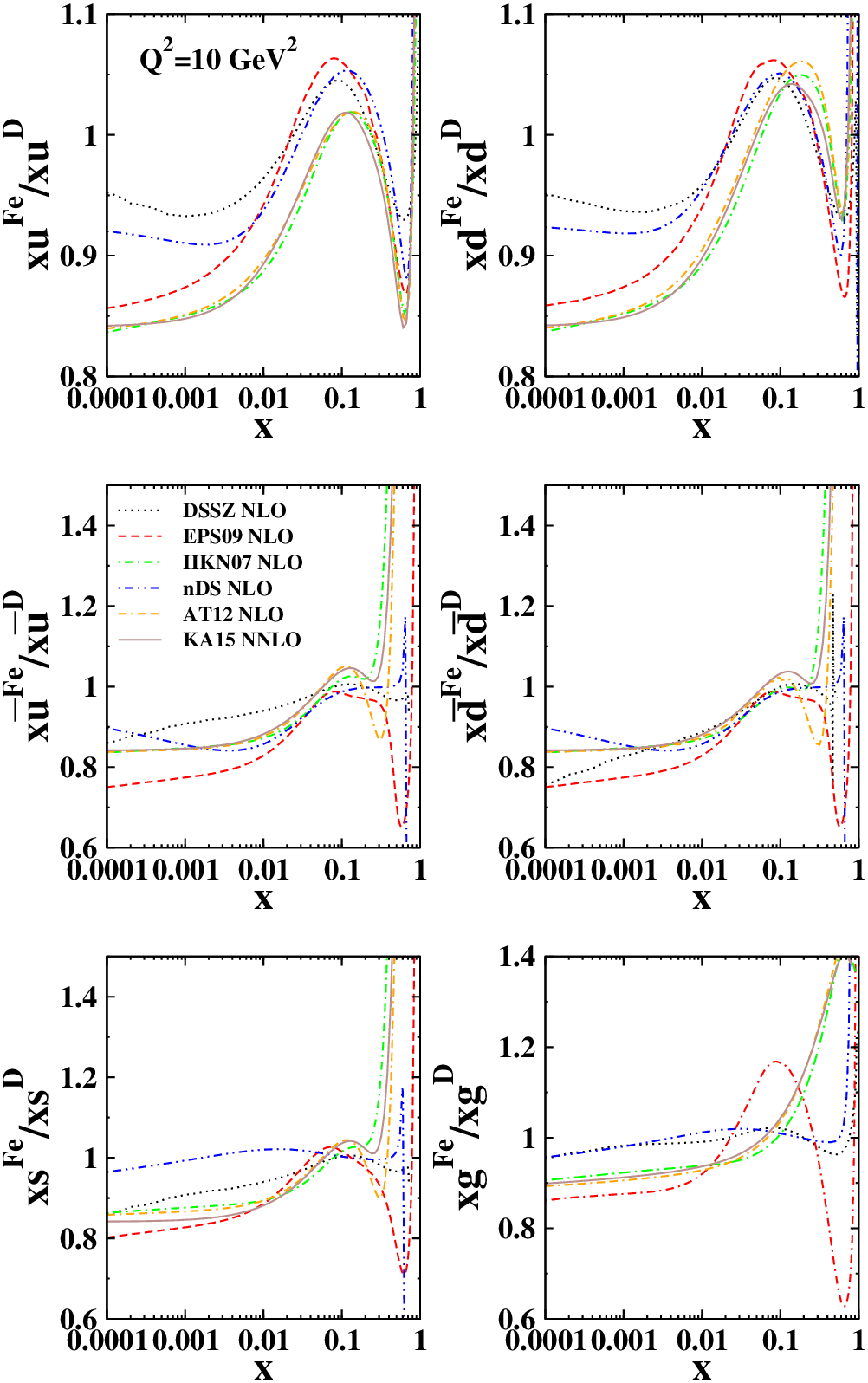}} 
		\caption{(Color online) The obtained NNLO nuclear modification factors, {\tt KA15}, defined in Eq.~\ref{weight} as a function of $x$ for iron at Q$^2$ = 10 GeV$^2$. The results from other groups such as {\tt AT12}\cite{AtashbarTehrani:2012xh}, {\tt EPS09} \cite{Eskola:2009uj}, {\tt HKN07}\cite{Hirai:2007sx}, {\tt nDS}\cite{deFlorian:2003qf} and {\tt DSSZ}\cite{deFlorian:2012qw} have been shown for comparisons.}\label{fig:ratiopartonFeQ10}
	\end{center}
\end{figure}

A detailed comparison of different approach resulting from the available nuclear PDFs analyses can be found in figure~\ref{fig:ratiopartonFeQ10}. The plots show that the differences are noticeable.
For almost all PDFs at Q$^2$ = 10 GeV$^2$ shown for Fe in the figure, our NLO and NNLO nuclear PDFs have significant overlap with {\tt HKN07} thorough much of the $x$ range. It is due to that the technical framework and data set selection of our global analysis are closest to {\tt HKN07} nuclear PDFs analysis. For the $u$ and $d$ PDFs, both our NLO and NNLO results including {\tt HKN07} show a stronger shadowing suppression at small values of $x$.
In medium to small $x$ of $\bar{u}$, $\bar{d}$ and $s$ PDFs, we have slight overlap with other nuclear PDFs sets.
For the gluon PDF, there is a variation among the different PDFs sets. The {\tt AT12}, {\tt HKN07}, {\tt EPS09} and {\tt KA15} $g$ PDFs all agree very nicely with each other throughout the medium to small $x$, ($0.001  \lesssim x \lesssim 0.1$). However the {\tt EPS09} shows stronger shadowing suppression at small values of $x$. It has also a larger enhancement in the anti-shadowing region (x $\sim$ 0.1). The {\tt DSSZ12} and {\tt nDS} gluon PDFs agree nicely throughout much of the $x$ range.

\begin{figure}[htb]
\begin{center}
\vspace{0.99cm}
\resizebox{0.45\textwidth}{!}{\includegraphics{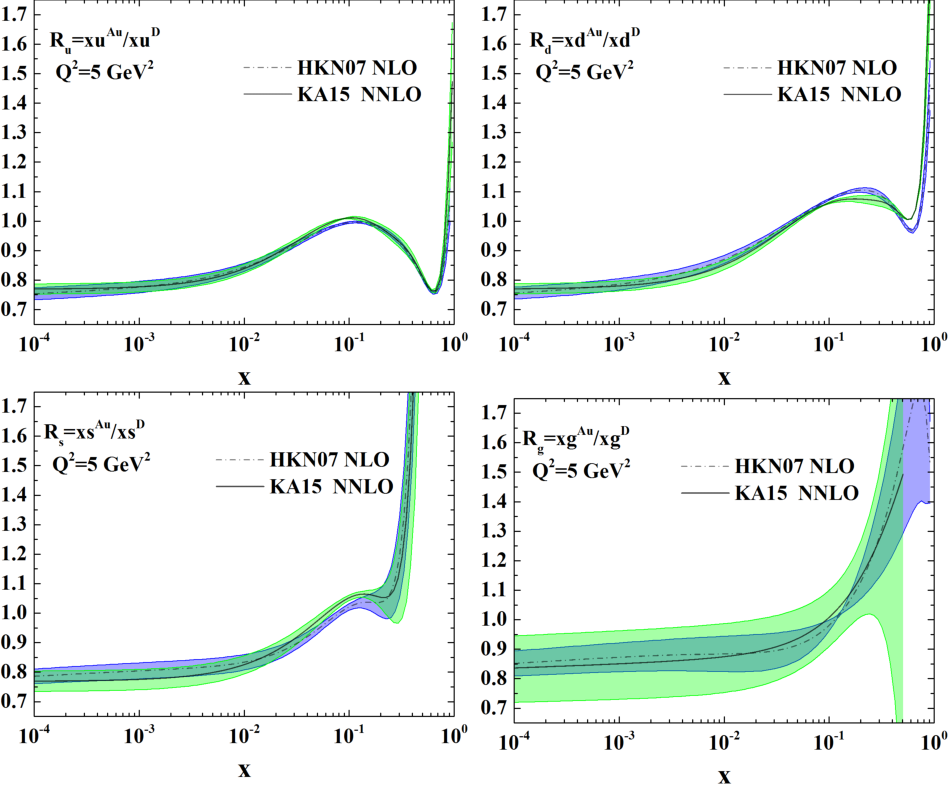}} 
\caption{(Color online) Comparison of the {\tt KA15} fit (green) with the results obtained by {\tt HKN07} (blue)~\cite{Hirai:2007sx}. The ratios are plotted as a function of $x$ at the scale Q$^2$=5 Ge$V^2$ for a gold nucleus. The error bands show the uncertainty of the nuclear PDFs.}\label{fig:figHKN07}
\end{center}
\end{figure}

\begin{figure}[htb]
	\begin{center}
		\vspace{0.99cm}
		\resizebox{0.45\textwidth}{!}{\includegraphics{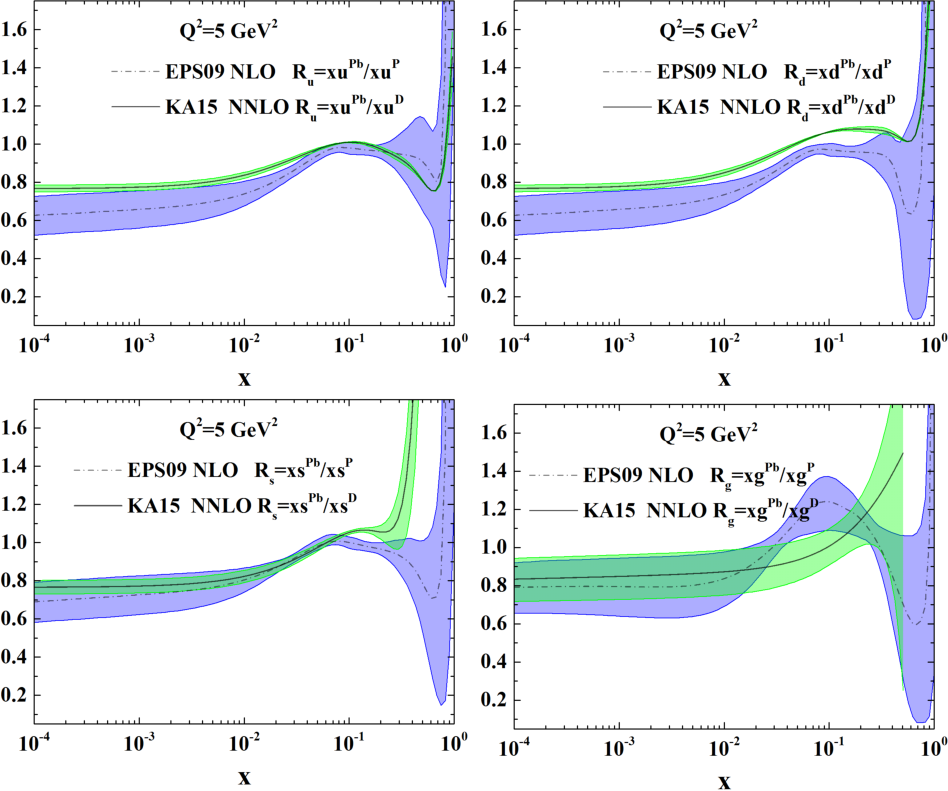}} 
		\caption{(Color online) Comparison of the {\tt KA15} fit (green) with the results obtained by {\tt EPS09} (blue)~\cite{Eskola:2009uj}. The ratios are plotted as a function of $x$ at the scale Q$^2$=5 Ge$V^2$ for a lead nucleus. The error bands show the uncertainty of the nuclear PDFs.}\label{fig:figEPS09}
	\end{center}
\end{figure}

In figure~\ref{fig:figHKN07} and~\ref{fig:figEPS09}, we plot nuclear modifications for the nuclear PDFs of a proton bound in gold and lead respectively. We show the results for these rather heavy nuclei, because they are the main targets at the heavy ion colliders such as LHC.  The ratios are plotted as a function of $x$ at the scale Q$^2$=5 Ge$V^2$. The results from  {\tt HKN07} and  {\tt EPS09} global nuclear PDFs analyses are also have been shown for comparison. 
For the $u$, $d$ and $s$ PDFs, we have overlap with {\tt HKN07} results while the nuclear gluon PDFs has larger shadowing suppression at small values of $x$ than {\tt HKN07} analysis.

In comparison with {\tt EPS09} which has been shown in the figure~\ref{fig:figEPS09}, the uncertainty bands for all nuclear PDFs of our NNLO analysis are considerably smaller than the uncertainty band of {\tt EPS09} throughout much of the $x$ range.

\begin{figure}[htb]
	\begin{center}
		\vspace{0.99cm}
		\resizebox{0.45\textwidth}{!}{\includegraphics{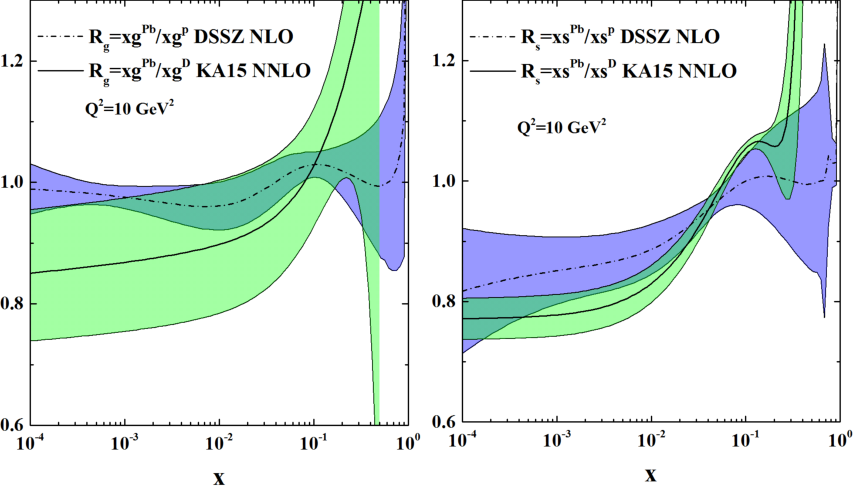}} 
		\caption{(Color online) Comparison of the {\tt KA15} fit (green) with the results obtained by {\tt DSSZ} (blue)~\cite{deFlorian:2011fp}. The ratios are plotted as a function of $x$ at the scale Q$^2$=10 Ge$V^2$ for the lead nuclei.  The error bands show the uncertainty of the nuclear PDFs.}\label{fig:DSSZ}
	\end{center}
\end{figure}

In figure\ref{fig:DSSZ}, we plot the $s$ and gluon PDFs as a function of $x$ at the scale Q$^2$=10 Ge$V^2$ for the lead nuclei. As we mentioned in Eq.~\ref{ssbar}, we relate the strange distribution to the light quarks sea ($\bar{u}$ and $\bar{d}$) distributions, consequently the $s$ distribution doesn't contribute in the fitting processes directly. As a result the strange distribution $s$ is similar to the light quarks sea distributions. This behavior leads to considerably smaller uncertainties for the strange distribution. The {\tt DSSZ12} assume that the light strange quark $s$ and anti-quark $\bar{s}$ have the same modification factors and relate them to the valance quarks modification factors. As the  figure~\ref{fig:DSSZ} shows, the {\tt DSSZ12} light strange quark distribution has bigger uncertainties both for small and large $x$. For the gluon distribution, this treatment is rather different. The {\tt DSSZ12} analysis shows a better description of EMC effect and also smaller uncertainty band. The gluon shadowing in the small $x$ region ($x  \lesssim 0.01$) has been constrained by the momentum sum rule and indirectly by the Q$^2$ evolution effects in the sea quarks sector which reflected by the DIS and Drell-Yan (DY) data. In addition, the inclusion of new and more precise measurements for example high-$p_T$ data from RHIC will provide important further constraints for the gluon shadowing region. The obtained gluon PDFs from {\tt KA15} analysis shows a stronger gluon shadowing at small $x$.

The mentioned differences between available nuclear PDFs analyses presented in this section generally arise from two source, the selection of data points used in the global analysis and direct parameterization of the nuclear PDFs or parameterization of nuclear modifications factors. Overall we found relatively good agreements between different nuclear PDFs sets.

%
%
\section{Nuclear PDFs at the LHC era} \label{LHC-era}

The proton+lead (p--Pb) and lead+lead (Pb--Pb) collisions are an integral part of the present and future nuclear programs at the Large Hadron
Collider (LHC).
As we mentioned, the nuclear PDFs are essential tools in high energy heavy--ion nucleus--nucleus (A--A) collisions at the future RHIC and CERN-LHC programs.
Generally speaking, the nuclear PDFs has very important role in the ongoing LHC proton+lead and lead+lead collisions. Some works have been
done in this regard to conclusively test the universality of the nuclear PDFs and also to investigate the sensitivity of the nuclear modifications in the  PDFs~\cite{Paukkunen:2014nqa,Armesto:2013kqa,Eskola:2013aya,Paukkunen:2014vha}.
The first experimental results published by the ALICE and CMS collaborations for the proton-lead (p--Pb) collisions at a nucleon--nucleon centre--of--mass energy of $\sqrt{s_{NN}}$ = 5.02 TeV  are summarized in details in Refs.~\cite{Loizides:2014oba,Chatrchyan:2014hqa}.
The CMS Collaboration also has recorded 150 $\mu b^{-1}$ in Pb--Pb collisions at $\sqrt{s_{NN}}$ = 2.76 TeV~\cite{GranierdeCassagnac:2014jha}.
All the heavy-ion public physics results from the CMS Collaboration are collected in Ref.~\cite{CMS-PbPb-data}. Some experimental studies on different aspects of heavy-ion collisions are presented in Refs~\cite{Salur:2015fza,Adam:2015xea,Paukkunen:2015bwa,Adam:2015hoa,Khachatryan:2015xaa,Kim:2014jia,DelaCruz:2014hva,Alvioli:2014eda,Barbieri:2014kaa,Lee:2014baa,Khachatryan:2015hha,Chatrchyan:2012nt,Zsigmond:2014wia,Ru:2014yma}.

\begin{figure}[htb]
	\begin{center}
		\vspace{0.95cm}
		\resizebox{0.45\textwidth}{!}{\includegraphics{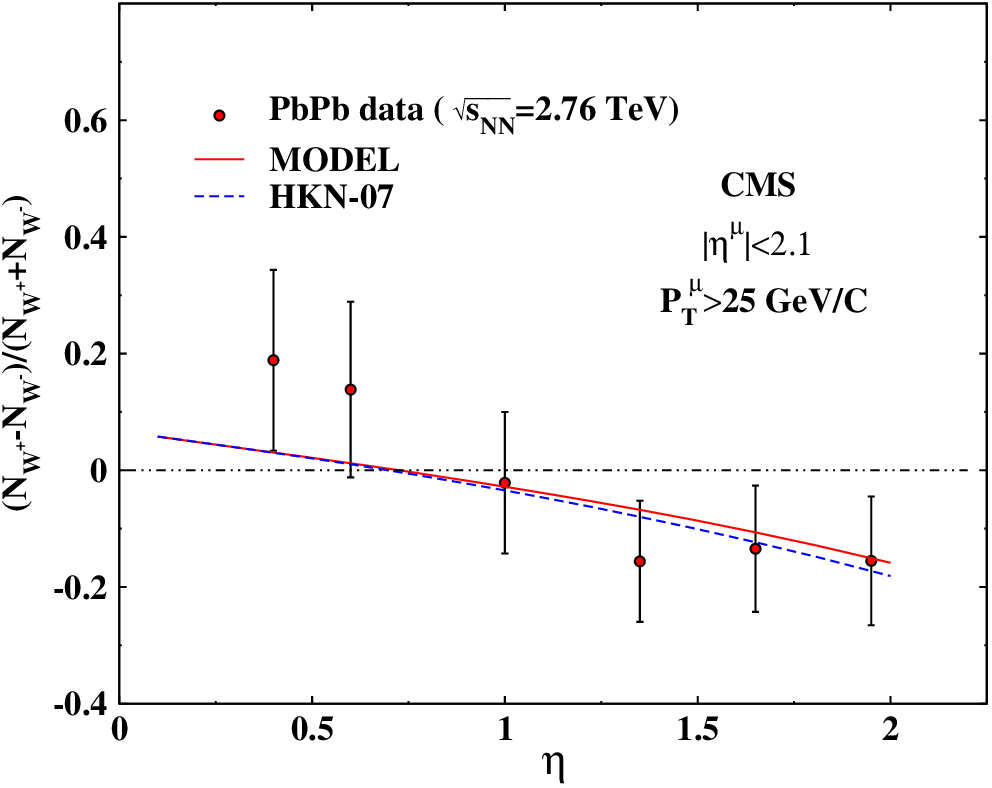}} 
		\caption{ (Color online) Lepton ($\ell = \mu$) charge asymmetry as a function of muon pseudo-rapidity for Pb--Pb collisions recorded by CMS at $\sqrt{s_{NN}}$ = 2.76 TeV~\cite{Chatrchyan:2012nt}. The red circles represent the data and the solid curve represent our theoretical predictions. }\label{fig:cms}
	\end{center}
\end{figure}

W and Z boson production in proton-nucleus (p--A) and nucleus-nucleus (A--A) collisions at the CERN-LHC offer a unique opportunity to probe nuclear PDFs.
The CMS collaboration at CERN presented their first study on W production (via leptonic decay channel) in Pb--Pb collisions at $\sqrt{s_{NN}}$ = 2.76 TeV~\cite{Chatrchyan:2012nt} and in p--Pb collisions at $\sqrt{s_{NN}}$ = 5.02 TeV~\cite{Khachatryan:2015hha} and also the Z boson production (via dimuon and dielectron decay channels) in Pb--Pb collisions at $\sqrt{s_{NN}}$ = 2.76 TeV~\cite{Chatrchyan:2014csa}. From discussion that we made in the paper, more data are needed to constrain the nuclear PDFs. The main difficulty of all global analysis of nuclear PDFs is the lack of any DIS data with heavy-ion beams which lead to larger uncertainties. For this reason, the obtained nuclear PDFs are less precisely known for nuclei than for the nucleons.
As a consequence, precise measurements of the W boson production in heavy-ion collisions and including the corresponding data in any global fits may lead to an improved determination of the nuclear PDFs~\cite{Khachatryan:2015hha}. Moreover, lepton charge asymmetry via dominant processes at the LHC in W$^+$ ($ u \bar{d} \rightarrow W^+ \rightarrow \ell^+ \nu_\ell$) and W$^-$ ($ \bar{u} d \rightarrow W^- \rightarrow \ell^- \nu_\ell$) productions, can permit the flavour asymmetries of $d$ and $u$ quark distributions in the nuclei.
The lepton (muon) charge asymmetry in Pb--Pb collisions collected by the CMS experiment at $\sqrt{s_{NN}}$ = 2.76 TeV~\cite{Chatrchyan:2012nt} is shown in figure.~\ref{fig:cms} and compared to our theoretical predictions. The theoretical results from {\tt HKN07}~\cite{Hirai:2007sx} are also shown as well.

In addition to the lepton charge asymmetry, study on the nuclear modification factor of the PDFs is the current interest in the recent heavy-ion collisions at CERN-LHC~\cite{Chatrchyan:2014csa,Armesto:2015lrg,Chatrchyan:2011ua,Zsigmond:2015wna,Zsigmond:2014wia}. The present measurements may set significant constraints for the global fits of the nuclear PDFs in an unexplored kinematical region of $x$.

Furthermore, recent detailed studies show the possibilities of direct measurements of large-mass elementary particles such as Higgs boson and top-quark via heavy-ion collision at the multi-TeV CERN-LHC and proposed future circular collider (FCC)~\cite{d'Enterria:2015jna,Baskakov:2015nxa}. Double-top or single-top productions in lead-lead (Pb--Pb) and proton-lead (p--Pb) collisions can be used to constrain the nuclear PDFs, specially the nuclear gluon distribution in small value of momentum fraction, $x \approx$ 10$^{-3}$--10$^{-2}$. Our study on the single and pair-production of top-quark at LHC and FCC energies via p--Pb and Pb--Pb collisions is in progress.

%
%
\section{Summary and conclusions} \label{Conclusions}

We presented, for the first time, a global analysis of nuclear PDFs and their uncertainties at the next-to-next-to-leading order (NNLO) accuracy in perturbative QCD.
We performed a $\chi^2$ analysis using available DIS $\ell ^\pm$ + nucleus  and Drell-Yan data.
The uncertainties of the determined nuclear PDFs are estimated using the Hessian method. The nuclear charm quarks distributions are also added into the analysis.
The result of the fit is a set of nuclear PDFs which incorporate the $x$, $Q$ and also A dependence,
so one can accommodate the full range of nuclear targets from light A = 2 to heavy A = 208.
A good fit to data has been obtained. We find a good agreement with experimental data and other fits.
As new and more precise measurements of observables sensitive to the gluon distribution will become available in the future high energy experiments,
we are expecting smaller uncertainty on the fitted nuclear PDFs.
In this respect, data from the CERN-LHC proton-lead run including dijet data from the CMS collaboration, are foreseen to bring significant additional insight.
It will also provide us a new window in theoretical understanding of the high energy process involving nucleon.
There are large amount of precise data for the free proton case, so one can develop a combined analysis of proton PDFs and nuclear PDFs.
The combination of PDFs and nuclear PDFs analysis provide good constraints on the gluon distributions at small values of Bjorken-$x$ and may allow for a good separation of the quark flavours in a wide range of $x$, which are mostly important for the present and future collider phenomenological tasks.
The new measurements of the nuclear effects in the Drell-Yan production which is planned in the E906/Drell-Yan experiment at Fermilab~\cite{E906-1,E906-2,E906-3} would be interested to analysis. The primary goals of this measurement at Fermilab include the study of the anti-$d$ to anti-$u$ quark asymmetry in the proton and a detailed study of the EMC effect in sea quarks.
Our next goal is to perform the present analysis, as well as, when the mentioned data become
finalized, including upcoming heavy--ion collisions data sets from CERN-LHC~\cite{d'Enterria:2015jna} and photon production in d + Au and Au + Au
from PHENIX~\cite{Cazaroto:2008qh,Tannenbaum:2007sy,Adler:2006wg}.
Selecting a complete data set plays a major role in constraining the nuclear modifications in any nuclear PDFs analysis. Further constraints for the nuclear gluon distributions in the yet unexplored regions of the $x$ and Q$^2$
plane are absolutely necessary for understanding QCD parton dynamics in
hadronic and nuclear high-energy collisions.
Our next-to-next-to-leading order of nuclear PDFs including their uncertainties can be calculated using the codes discussed in Appendix.{\bf B}.

%
%
\section*{Appendix A: Sum rules, baryon number and momentum conservation}\label{AppendixA}

Using three sum rules presented in Eq.~\ref{Constraints} which give us the nuclear charge $Z$, baryon number $A$ and momentum conservation,
one can calculate the three parameters $a_{u_v}(A,Z)$, $a_{d_v}(A,Z)$ and $a_g(A,Z)$. For practical usage, we
express these constants by eight integral values $I_1$--$I_8$ as we explained in our previous version of nuclear PDFs~\cite{AtashbarTehrani:2012xh}:

\begin{eqnarray}
a_{u_v}(A,Z)& = & -\frac{Z I_1(A) + (A - Z) I_2(A)}{Z I_3 + (A - Z) I_4 }, \nonumber \, \, \, \, \, (A1) \\
a_{d_v}(A,Z)& = & -\frac{Z I_2(A) + (A - Z) I_1(A)}{Z I_4 + (A - Z) I_3 }, \nonumber \\
a_g(A,Z)& = & -\frac{1}{I_8}\left\{a_{u_v}(A,Z)\left[ \frac{Z}{A} I_5 + \left(1-\frac{Z}{A}\right) I_6\right]
\right.\nonumber\\
& + & \left. a_{d_v}(A,Z)\left[\frac{Z}{A} I_6 + \left(1 - \frac{Z}{A} I_5\right)\right] + I_7(A)\right\} \,. \nonumber
\label{au,ad,ag}
\end{eqnarray}\label{eq:a}
The numerical values of the eight integrals are listed in Table~\ref{table5} from the present analysis for Lead.
\begin{table}[htb]
\begin{center}
\begin{tabular}{cccc}
\hline\hline
Integral & Value & Integral & Value \\ \hline\hline
I$_1$ & 0.0890676 & \multicolumn{1} {|c}{I$_5$} & 0.374181 \\
I$_2$ & 0.0537472 & \multicolumn{1} {|c}{I$_6$} & 0.156111 \\
I$_3$ & 2.1693    & \multicolumn{1} {|c}{I$_7$} & 0.0226629 \\
I$_4$ & 1.06856   & \multicolumn{1} {|c}{I$_8$} & 0.39874 \\ \hline\hline
\end{tabular}
\caption{ Numerical values of the eight integrals for Lead. }\label{table5}
\end{center}
\end{table}

\begin{figure}[htb]
\begin{center}
\vspace{1.1cm}
\resizebox{0.35\textwidth}{!}{\includegraphics{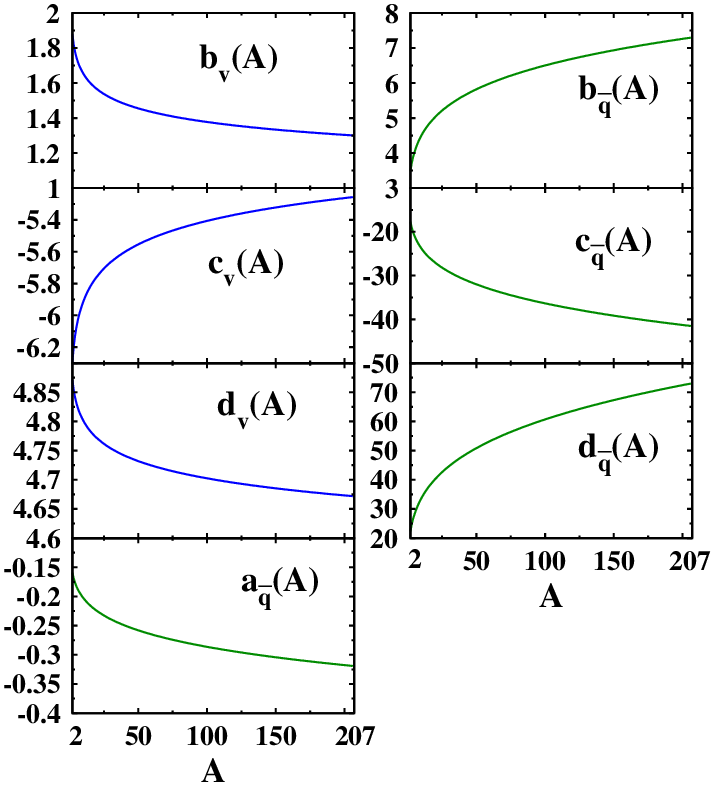}} 
\caption{ (Color online) A-dependence of the fit parameters according to Eq.~(\ref{weight}).}\label{fig:cof}
\end{center}
\end{figure}

Using these values together with Eq.(A1), one could calculate the constants:
\newline
$a_{u_v}$ = -0.0450391, $a_{d_v}$ = -0.0433013 and $a_g$ = 0.00208932,
\newline
for Lead. The A-dependence of the parameters in Eq.~(\ref{weight}) are plotted in figure.~\ref{fig:cof}.

%
%
\section*{Appendix B: FORTRAN package of {\tt KA15} nuclear PDFs}\label{AppendixB}

We prepared a code for calculating the nuclear PDFs including their uncertainties at different values of $x$ and Q$^2$. The \texttt{FORTRAN} package containing our unpolarized structure functions, $F_{2}^{(A,Z)}(x,Q^{2})$, for nuclei as well as the
nuclear parton densities $xu_{v}^A(x,Q^{2})$, $xd_{v}^A(x,Q^{2})$, $x\bar{u}^A(x,Q^{2})$,
$x\bar{d}^A(x,Q^{2})$, $xs^A(x,Q^{2})$, $xc^A(x,Q^{2})$, $xg^A(x,Q^{2})$ and their uncertainties at NNLO approximation
in the $\overline{{\rm MS}}$--scheme can be obtained via e-mail from the authors.
In this package we assumed the following kinematical ranges $10^{-4} \leq x \leq 0.999$ and $1 \leq Q^2 \leq 10^{5}$ GeV$^2$ for the $x$ and Q$^2$ respectively. The obtained nuclear PDFs can be used for high-energy nuclear reactions to study the nuclear effects.

%
%
\section*{Acknowledgments}
The authors are especially grateful to Fredrick Olness from SMU for the
fruitful discussions and critical remarks. We also would like to thank Mojtaba Mohammadi, Abolfazl Mirjalili and Muhammad Goharipour for carefully reading the manuscript and their helpful comments. We are also grateful to Rodolfo Sassot and Pia Zurita for providing us with the best fit of {\tt DSSZ12} nuclear PDFs. Authors are thankful School of Particles and Accelerators, Institute for Research in Fundamental Sciences (IPM) for financially support of this project.
H. Khanpour also thanks the University of Science and Technology of Mazandaran for financial support provided for this research.


%
%

\end{document}